\documentclass{article}

\usepackage{microtype}
\usepackage{graphicx}
\usepackage{subcaption}
\usepackage{booktabs} 

\usepackage{hyperref}

\usepackage[preprint]{icml2026}

\usepackage{amsmath}
\usepackage{amssymb}
\usepackage{mathtools}
\usepackage{amsthm}

\usepackage[table]{xcolor}
\usepackage{colortbl}
\usepackage{multirow}

\usepackage[capitalize,noabbrev]{cleveref}

\theoremstyle{plain}

\theoremstyle{definition}

\theoremstyle{remark}

\usepackage[textsize=tiny]{todonotes}

\icmltitlerunning{DLM: Unified Decision Language Models for Offline Multi-Agent Sequential Decision Making}

\begin{document}

\twocolumn[
  \icmltitle{DLM: Unified Decision Language Models for Offline Multi-Agent Sequential Decision Making}



  \icmlsetsymbol{equal}{*}

  \begin{icmlauthorlist}
    \icmlauthor{Zhuohui Zhang}{tongji-cse,shanghai-institutes}
    \icmlauthor{Bin Cheng}{tongji-cse,shanghai-institutes}
    \icmlauthor{Bin He}{tongji-cse,shanghai-institutes}
  \end{icmlauthorlist}

  \icmlaffiliation{tongji-cse}{Department of Control Science and Engineering, Tongji University, Shanghai 201804, China}
  \icmlaffiliation{shanghai-institutes}{Shanghai Institute of Intelligent Science and Technology; Shanghai Research Institute for Intelligent Autonomous Systems; State Key Laboratory of Autonomous Intelligent Unmanned Systems, Shanghai 200120, China}

  \icmlcorrespondingauthor{Bin Cheng}{bincheng@tongji.edu.cn}

  \icmlkeywords{Machine Learning, ICML}

  \vskip 0.3in
]



\printAffiliationsAndNotice{}  

\begin{abstract} 
  Building scalable and reusable multi-agent decision policies from offline datasets remains a challenge in offline multi-agent reinforcement learning (MARL), as existing methods often rely on fixed observation formats and action spaces that limit generalization. In contrast, large language models (LLMs) offer a flexible modeling interface that can naturally accommodate heterogeneous observations and actions. Motivated by this, we propose the Decision Language Model (DLM), which formulates multi-agent decision making as a dialogue-style sequence prediction problem under the centralized training with decentralized execution paradigm. DLM is trained in two stages: a supervised fine-tuning phase, which leverages dialogue-style datasets for centralized training with inter-agent context and generates executable actions from offline trajectories, followed by a group relative policy optimization phase to enhance robustness to out-of-distribution actions through lightweight reward functions. Experiments on multiple benchmarks show that a unified DLM outperforms strong offline MARL baselines and LLM-based conversational decision-making methods, while demonstrating strong zero-shot generalization to unseen scenarios across tasks.
\end{abstract}

\section{Introduction}
\label{s_1}

Large language models (LLMs)~\citep{touvron2023llama, bai2023qwen, ouyang2022training, brown2020language}, trained on offline datasets, have demonstrated generalization in a wide range of downstream tasks in natural language processing~\citep{guo2024deepseek, schick2023toolformer}. Nevertheless, LLMs fall short when applied to sequential decision-making problems due to misalignment with task goals and environment dynamics~\citep{ahn2022can}, which are better handled by reinforcement learning (RL). In contrast, online RL~\citep{sutton1998reinforcement, zhang2025bridging, schulman2017proximal, haarnoja2018soft} relies on repeated environment interaction, which is expensive, inefficient, and risky in real-world applications. As a result, offline RL~\citep{kumar2020conservative, fujimoto2019off, kostrikov2021offline} has emerged as a promising alternative that learns from static datasets without additional interaction. However, most existing offline RL methods focus on improving single-task performance under fixed datasets by mitigating out-of-distribution (OOD) issues~\citep{kumar2019stabilizing, zhang2025pagnet, levine2020offline}, limiting their ability to generalize across diverse tasks. At the same time, applications such as autonomous driving~\citep{zhou2024behaviorgpt}, collaborative robotics~\citep{seraj2023mixed}, and strategic games~\citep{rashid2020weighted, mixrts} demand both generalization across various tasks and scalability with increasing numbers of interacting agents. Addressing these challenges requires a unified model that can handle multi-agent sequential decision-making across tasks within a single framework.

A reason for the limited generalization in RL lies in the rigid construction of models. States and actions are encoded in fixed formats that are tightly coupled with task definitions, hindering transferability between environments with different input and output structures. Recent efforts such as the decision transformer (DT)~\citep{chen2021decision} and the trajectory transformer (TT)~\citep{janner2021offline} address this limitation in single agent settings by casting decision-making as sequence modeling, allowing a more flexible and data-driven formulation. However, extending these methods to multi-task scenarios introduces a representation mismatch challenge. Most existing approaches perform modality-specific embeddings by normalizing and discretizing continuous inputs into bounded index representations. This strategy struggles to generalize across heterogeneous scenarios due to large variations in input distributions, resulting in unstable and scenario-dependent representations. The situation becomes more complex in multi-agent settings, which introduce additional coordination and scalability challenges. One challenge is incompatibility of centralized training with decentralized execution (CTDE)~\citep{lowe2017multi}. While the CTDE paradigm aims to leverage global information during training and rely on local observations at execution, reconciling these two modes within a single model remains difficult, as it requires balancing global context with agent-level autonomy. Another challenge arises from the sparsity and delayed nature of reward signals in multi-agent environments, where rewards are episodic or only available at the trajectory level, exacerbating the credit assignment problem for sequence modeling approaches.

In this paper, we propose the Decision Language Model (DLM), a scalable framework designed to alleviate core challenges in multi-agent sequential decision-making across heterogeneous scenarios, while reducing out-of-distribution (OOD) errors under limited data. DLM is trained in two stages: a supervised fine-tuning (SFT) phase, which adapts a pre-trained language model to the decision-making domain (DLM-SFT), followed by a group relative policy optimization (GRPO)~\citep{shao2024deepseekmath} phase that improves robustness to OOD actions (DLM-GRPO). To tackle the representation mismatch, we convert observations and actions into natural language and reformulate decision-making as dialogue-style sequence modeling, allowing LLMs to encode diverse tasks through tokenization. To address the CTDE incompatibility, we design a dialogue-style trajectory representation for each agent, which preserves inter-agent context during centralized training while supporting decentralized execution from local observations. To cope with sparse and delayed reward signals, we leverage return-to-go values from offline trajectories to construct a return-aware reward signal in the GRPO stage, enabling relative policy optimization, while incorporating an executability constraint to penalize invalid actions and reduce OOD risks.

We evaluate DLM across cooperative multi-agent benchmarks, including level-based foraging (LBF)~\cite{rangwala2020learning}, starcraft multi-agent challenge (SMAC)~\citep{samvelyan2019starcraft}, and SMACv2~\citep{ellis2024smacv2}, covering a diverse set of tasks and scenarios, and compare it against value-based offline MARL methods and LLM-based methods. The results show that DLM-SFT, trained from offline observation-action trajectories, achieves performance comparable to offline MARL baselines. Building on this, DLM-GRPO further improves performance, showing better decision quality and generalization compared to LLM-based methods in multi-agent settings. Overall, a unified DLM can handle diverse multi-agent sequential decision-making problems across tasks.

\begin{itemize}
\item We propose the DLM, a framework for multi-agent sequential decision-making across tasks, trained via a two-stage offline pipeline that addresses representation mismatch, CTDE incompatibility, and sparse reward challenges without relying on online interaction.
\item We introduce a dialogue-style offline dataset construction paradigm for MARL and an LLM-based post-training approach aligned with CTDE, enabling centralized training with inter-agent context and decentralized execution from local observations.
\item Through extensive experiments on multiple benchmarks, we show that a single unified DLM achieves performance competitive with or exceeding strong offline MARL and LLM-based conversational decision-making baselines, while demonstrating robust zero-shot generalization to unseen tasks.
\end{itemize}

\section{Related Works}
\label{s_2}
\begin{figure*}[t]
  \centering
  \centerline{\includegraphics[width=0.85\textwidth]{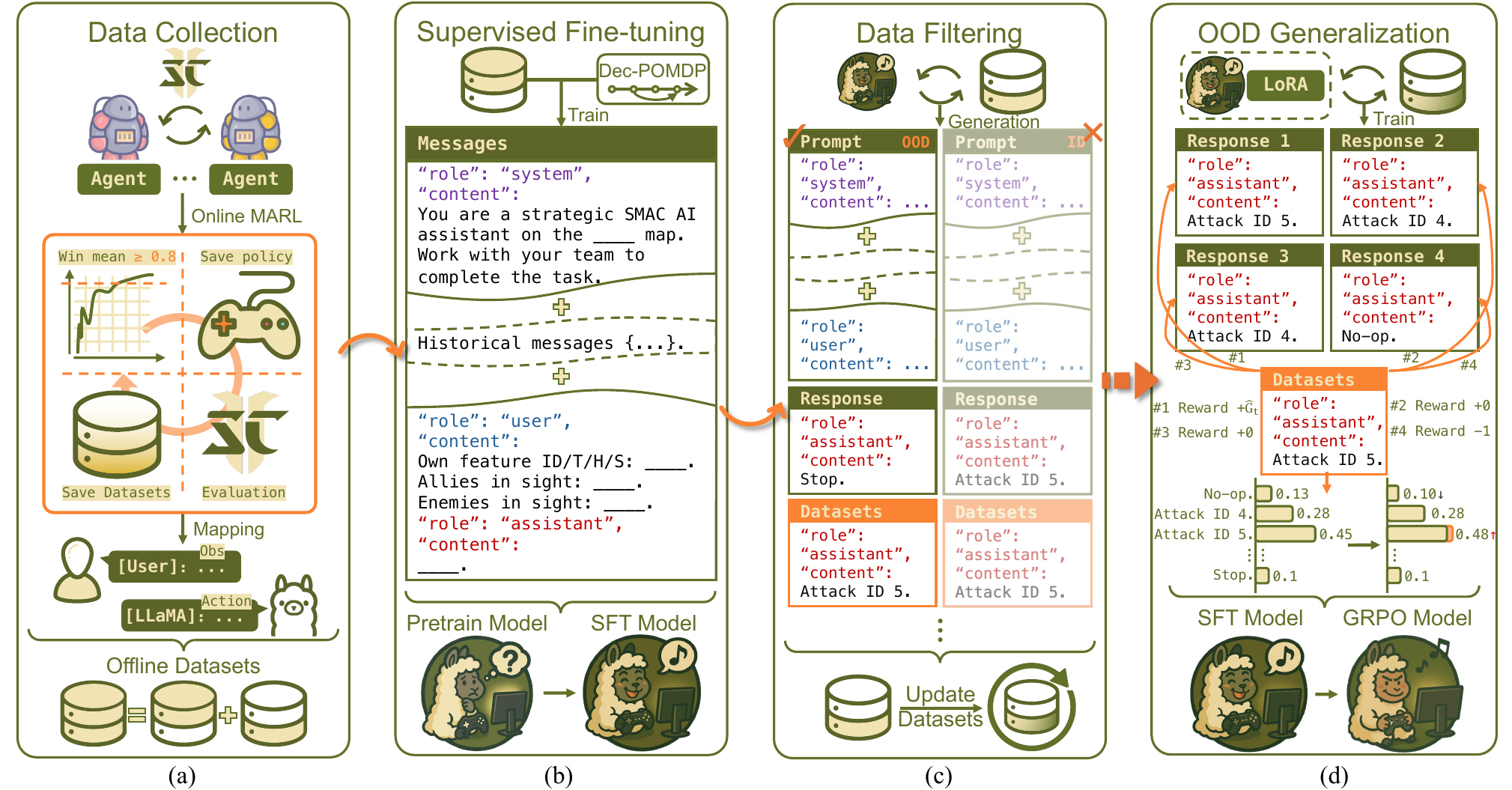}}
  \caption{Training pipeline of DLM. (a) Offline data collected from online MARL algorithms is transformed into dialogue-style sequences and split into two subsets. (b) The pre-trained model is fine-tuned on the first subset via SFT to align with the decision domain, resulting in DLM-SFT. (c) DLM-SFT generates policies on the second subset, and OOD-prone samples are filtered by comparing outputs with the dataset. (d) GRPO further trains DLM-SFT on the filtered subset using executability-based rewards, yielding the final DLM-GRPO model.}
  \label{f_1}
\end{figure*}
\paragraph{Offline MARL} Offline RL aims to learn policies from fixed datasets without further environment interaction, making it suitable for high-risk or cost-sensitive domains. A common approach is to apply behavior cloning (BC)~\citep{syed2008apprenticeship}, which directly imitates actions in the dataset. While simple and stable, BC does not account for the distributional shift between the training data and the learned policy’s behavior, often resulting in compounding errors during deployment. To address this, methods such as TD3+BC~\citep{kostrikov2021offline} and CQL~\citep{kumar2020conservative} introduce value-based regularization to penalize unseen or high-risk actions and reduce overestimation bias, achieving better performance in the single-agent setting. In the multi-agent setting, these issues become even more severe due to the exponential growth of the joint state-action space and the need for coordinated behavior. MACQL~\citep{formanek2024dispelling} extends CQL to the multi-agent regime by applying conservative value estimation under CTDE paradigm. OMIGA~\citep{wang2024offline} improves coordination by shaping local policies with implicit global-to-local value information. CFCQL~\citep{shao2024counterfactual} further enhances robustness through agent-level counterfactual regularization, enabling stable learning even under partial observability or suboptimal data coverage. Despite these advances, existing offline MARL methods are tied to task-specific architectures or value-centric learning objectives, 
limiting their scalability across diverse agents and environments.


\paragraph{Sequence Modeling for Decision Making} Recent advances have recast RL as a sequence modeling problem, enabling the use of Transformer~\citep{vaswani2017attention} architectures developed for language understanding. The DT~\citep{chen2021decision} frames policy learning as supervised sequence prediction by autoregressively predicting actions from past states and actions. Building on this, the TT~\citep{janner2021offline} models trajectory distributions with tokenized inputs, improving sample efficiency. Gato~\citep{reed2022generalist} unifies vision, language, and control tasks through sequence modeling, showing that a single transformer can operate across domains. SayCan~\citep{ahn2022can} demonstrates that LLMs can be used for high-level planning within a sequence modeling formulation. In the multi-agent setting, sequence modeling remains limited. Multi-agent decision transformer (MADT)~\citep{meng2023offline} shares parameters across agents to enable independent pre-training, and applies online fine-tuning with a centralized critic. While this design simplifies training, it neglects inter-agent coordination during pre-training and relies on environment interaction, deviating from the offline paradigm.

\paragraph{Aligning Pre-trained LLMs for Decision-Making} Pre-trained LLMs provide strong priors that help agents make informed decisions with minimal exploration, making them attractive for offline decision-making. However, these priors are often misaligned with target tasks or environments, motivating adaptation through alignment techniques. SFT, together with parameter-efficient methods such as LoRA~\citep{hu2021lora}, enables scalable adaptation in multi-agent settings. Beyond SFT, reinforcement learning from human feedback (RLHF)~\citep{ouyang2022training} further refines model behavior using preference-based rewards, yielding more robust policies. More recently, GRPO~\citep{shao2024deepseekmath} has simplified alignment by leveraging lightweight heuristic rewards, making it particularly suitable for offline RL where explicit reward design is costly or infeasible.

\section{Method}
\label{s_3}

In this section, we present DLM, a scalable framework for multi-agent sequential decision-making across tasks. As illustrated in Fig.~\ref{f_1}, DLM training follows a four-step pipeline consisting of two data preparation stages and two model training stages. Due to the lack of suitable offline datasets covering diverse tasks, we first construct a comprehensive offline multi-task dataset and transform the collected trajectories into a dialogue-style sequence representation. The dataset is then partitioned into two subsets for SFT and GRPO, respectively. We initialize DLM with a pre-trained LLaMA-3.2-1B model~\citep{grattafiori2024llama} and fine-tune it via SFT to obtain DLM-SFT, enabling the model to generate valid and rule-compliant actions from offline trajectories. Building on this, we apply GRPO with return-to-go signals for return-aware policy optimization, together with an executability constraint that penalizes invalid actions and improves robustness under distributional shifts.

\paragraph{Problem Formulation}

We model the cooperative multi-agent sequential decision-making problem as a decentralized partially observable markov decision process (Dec-POMDP)~\citep{oliehoek2016concise}, defined by the tuple $\mathcal{G} = \langle \mathcal{N}, \mathcal{S}, \mathcal{A}, P, \Omega, O, R, \gamma \rangle$.
Here, $\mathcal{N} = \{1, \dots, n\}$ denotes the set of agents, $\mathcal{S}$ is the set of global states, and $\mathcal{A} = \prod_{i=1}^n \mathcal{A}^i$ is the joint action space, where $\mathcal{A}^i$ is the action space for agent $i$.
At each time step $t$, the environment is in state $s_t \in \mathcal{S}$, and each agent $i$ receives a private observation $o_t^i \in \Omega^i$, where $\Omega = \prod_{i=1}^n \Omega^i$ is the joint observation space and $O: \mathcal{S} \to \Delta(\Omega)$ is the observation function.
Based on its local observation, each agent selects an action according to its individual policy $\pi^i: \Omega^i \to \Delta(\mathcal{A}^i)$.
The joint action $\boldsymbol{a}_t = (a_t^1, \dots, a_t^n)$ induces a transition to the next state $s_{t+1}$ according to the environment dynamics $P: \mathcal{S} \times \mathcal{A} \to \Delta(\mathcal{S})$. The system receives a global reward $r_t = R(s_t, \boldsymbol{a}_t)$, and $\gamma \in (0,1)$ denotes the discount factor governing future returns. In the offline setting, we assume access to a fixed dataset $\mathcal{D} = \{\tau^{(k)}\}_{k=1}^M$ consisting of $M$ collected trajectories. Each trajectory $\tau^{(k)}$ contains the key elements defined in the Dec-POMDP tuple $\mathcal{G}$. The goal is to learn decentralized policies $\{\pi^i\}_{i=1}^n$ to achieve cooperative behavior across diverse multi-agent tasks.

\subsection{Dialogue-Style Offline Dataset Construction}
\label{s_3_1}

\begin{figure}
  \centering
  \centerline{\includegraphics[width=1\columnwidth]{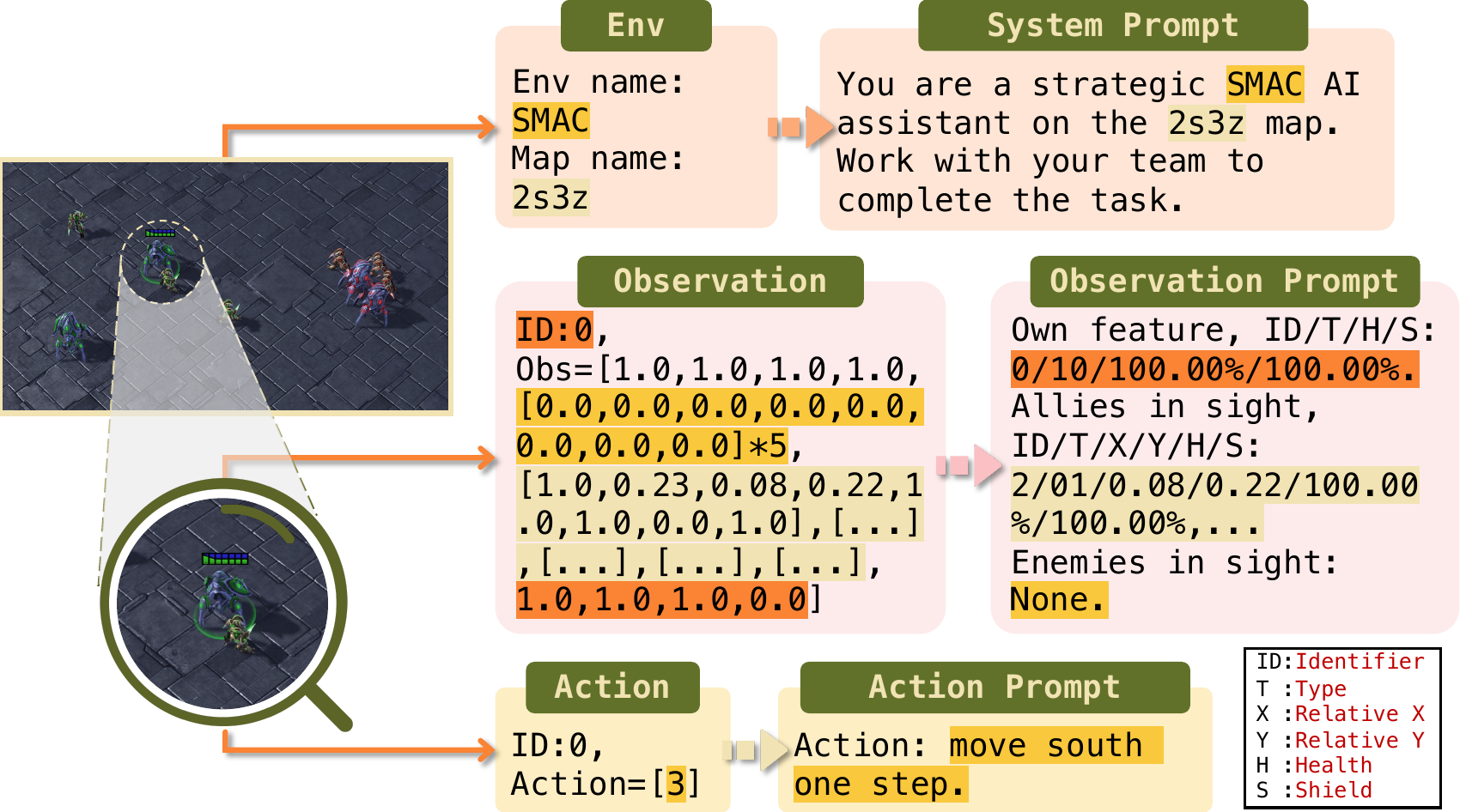}}
  \caption{Mapping SMAC environment, observations, and actions into a dialogue-style prompt, with highlighted text showing key correspondences.}
  \label{f_2}
\end{figure}

To address the representation mismatch challenge discussed in Sec.~\ref{s_1}, we rethink the encoding of observations and actions in multi-task settings. The wide variability in their numerical ranges and structures across tasks makes fixed-format mappings impractical. In contrast, LLMs leverage tokenization techniques~\citep{mikolov2013distributed} to embed diverse concepts into a shared semantic space, enabling flexible representation learning.

Inspired by this, we verbalize multi-agent decision trajectories as natural language dialogues, framing sequential decision-making as a language modeling problem. Any decision process can be described by specifying the environment, the agent's observation, and the intended action in natural language. Specifically, taking SMAC as an example, observations consist of four feature groups: \texttt{move\_feats}, \texttt{enemy\_feats}, \texttt{ally\_feats}, and \texttt{feats}, encoding attributes such as position, health, and visibility. These features, originally designed for computational processing, can be naturally verbalized into textual descriptions. As illustrated in Fig.~\ref{f_2}, environment information, agent observations, and actions are mapped into a fixed dialogue format, where observations are treated as the \texttt{user} input and actions as the \texttt{assistant} reply, forming a dialogue turn. We further convert the dialogue into the LLaMA-3 chat format for compatibility, as detailed in Fig.~\ref{f_A}. As existing datasets lack full coverage of SMAC tasks, we collect offline data following the dataset construction methodology used in D4RL~\citep{fu2020d4rl}. Specifically, we train TGCNet~\citep{zhang2025bridging} until the win rate exceeds 80\%, after which we save the model and use it to interact with the environment to generate trajectories. Details on dataset construction are provided in Appendix~\ref{s_A_1}.

\subsection{SFT for Multi-Agent Sequential Decision}
\label{s_3_2}
\begin{figure*}
  \centering
  \centerline{\includegraphics[width=0.85\textwidth]{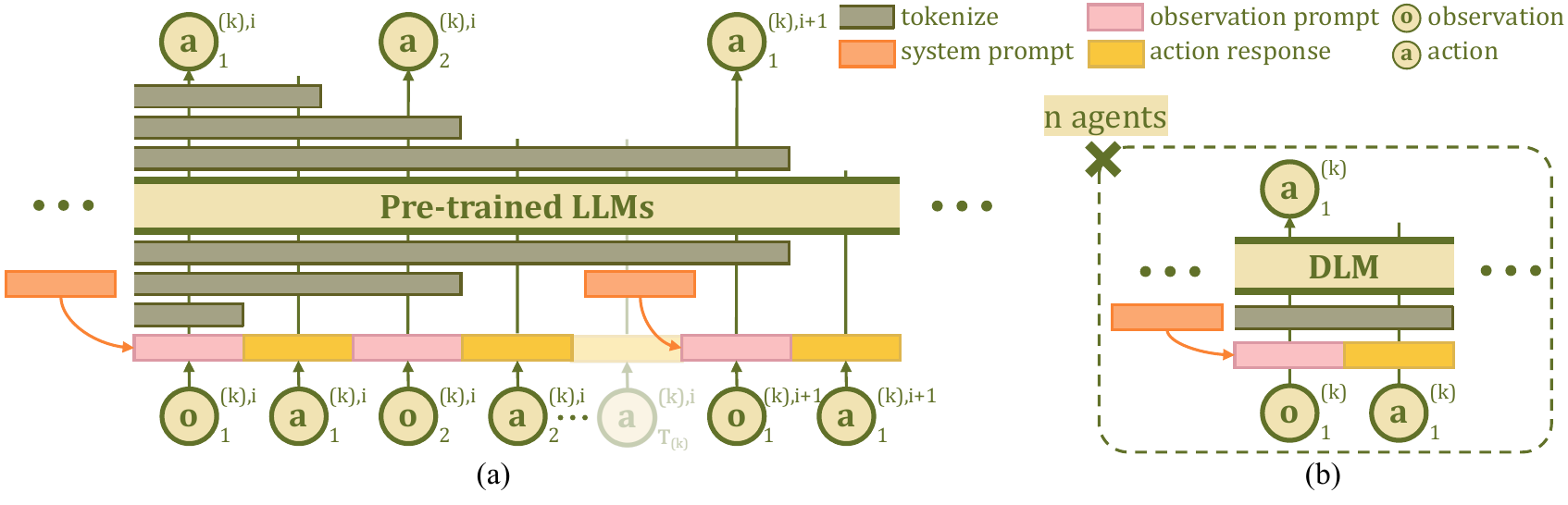}}
  \caption{Training and inference frameworks for DLM. (a) Centralized training using dialogue-style trajectories with inter-agent information. (b) Decentralized inference where each agent independently generates actions based on its local trajectory.}
  \label{f_3}
\end{figure*}

We adopt an autoregressive sequence modeling approach without relying on explicit value functions. Unlike prior single-agent formulations, we design a dialogue-style sequence representation that supports CTDE. Instead of pre-training from scratch, we initialize DLM with pre-trained LLMs via SFT on half of the constructed offline dataset $\mathcal{D}_{\text{SFT}}$. The overall training and inference framework is presented in Fig.~\ref{f_3}, with the specific SFT training procedure outlined in Alg.~\ref{a_1} and the inference procedure in Appendix Alg.~\ref{a_3}. The mathematical analysis and motivation behind this design are provided in the Appendix~\ref{app:design_motivation}.

\paragraph{Dialogue-Style Sequence Representation} 

We represent multi-agent decision trajectories as structured sequences of observation-action pairs. Formally, each trajectory $\tau^{(k)}$ is defined as:

\begin{equation}
\label{e_1}
\tau^{(k)} = \left( \, \left\{ (o_t^{(k),i},\ a_t^{(k),i})\ \big|\ t = 1,\ldots, T^{(k)} \right\} \, \right)_{i=1}^{N},
\end{equation}

where $N$ is the number of agents and $T^{(k)}$ denotes the length of the $k$-th trajectory. Here, $o_t^{(k),i}$ represents the observation of agent $i$ at time $t$ within the $k$-th trajectory, and $a_t^{(k),i}$ denotes the corresponding action. Since each observation-action pair has already been verbalized into a dialogue-style format, the model inputs are constructed by stacking them in the order specified in Eq.~\eqref{e_1}. To improve scalability for the number of agents, we apply a maximum token limit, truncating sequences that exceed it and adopting dynamic packing strategies.

\paragraph{SFT Training} 

We fine-tune DLM on the offline dataset, aligning the model's outputs with multi-agent decision demonstrations. In practice, DLM is trained to predict the next assistant reply (i.e., the action) conditioned on the dialogue history up to the current observation. Formally, the SFT objective minimizes the following loss:

\begin{equation}
\label{e_2}
\mathcal{L}_{\text{SFT}} = - \frac{1}{M} \sum_{k=1}^{M} \sum_{t=1}^{T^{(k)}} \sum_{i=1}^{N} \log P_\theta \left( a_t^{(k),i} \mid \tau_{\le (o_t^{(k),i})}^{(k)} \right),
\end{equation}

where $M$ is the number of trajectories, $T^{(k)}$ is the length of the $k$-th trajectory, $N$ is the number of agents, and $\tau_{\le (o_t^{(k),i})}^{(k)}$ denotes the dialogue history up to and including the current observation $o_t^{(k),i}$. 

\paragraph{Decentralized Inference}

During inference, DLM enables decentralized decision-making while maintaining the benefits of centralized training. Each agent independently generates its action based on its dialogue history, without requiring access to other agents' observations or actions. Formally, for each agent $i$ at time step $t$, the model predicts the next action by:
\begin{equation}
\label{e_3}
a_t^i = \arg\max_{a} P_{\theta_{\text{SFT}}}(a \mid \tau_{\le (o_t^i)}) .
\end{equation}
If the predicted action is invalid or not allowed by the environment's available actions, we resample by drawing from the predicted distribution, sampling an action from the truncated distribution where only the top-$k$ tokens whose cumulative probability exceeds the top-$p$ threshold are considered:
\begin{equation}
\label{e_4}
a_t^i \sim P_{\theta_{\text{SFT}}}^{\text{top-}p,\text{top-}k}\left( \cdot \mid \tau_{\le (o_t^i)} \right).
\end{equation}

\begin{algorithm}[tb]
\caption{SFT Training Procedure for DLM}
\label{a_1}
\begin{algorithmic}[1]
\STATE \textbf{Input:} Offline dataset $\mathcal{D}_{\text{SFT}}$, pre-trained parameters $\theta_{\text{init}}$, max length $L$
\STATE \textbf{Initialize:} $\theta_{\text{SFT}} \leftarrow \theta_{\text{init}}$, tokenizer, prompts
\FOR{each trajectory $\tau^{(k)} \in \mathcal{D}_{\text{SFT}}$}
  \STATE Tokenize $\tau^{(k)}$ by stacking $(o_t^{(k),i}, a_t^{(k),i})$ with system prompts, truncated to $L$
\ENDFOR
\STATE Construct mini-batches with dynamic packing
\FOR{each mini-batch}
  \STATE Predict next action conditioned on dialogue history
  \STATE Update $\theta_{\text{SFT}}$ by minimizing Eq.~\eqref{e_2}
\ENDFOR
\STATE \textbf{Output:} Fine-tuned model $\theta_{\text{SFT}}$
\end{algorithmic}
\end{algorithm}

\subsection{Filtering OOD Samples}
\label{s_3_3}

Although DLM-SFT learns reasonable decision behaviors from offline data, it occasionally generates invalid or OOD actions due to the inherent limitations of dataset coverage. In multi-agent settings like SMAC, the observation-action space is vast and continuous, making exhaustive offline coverage impractical. As illustrated by the t-SNE visualization in Appendix Fig.~\ref{f_4}, even with diverse trajectory collection across multiple tasks, the sampled observations and actions still only occupy a sparse subset of the overall space. This reveals a fundamental limitation: simply enlarging the dataset cannot completely eliminate OOD issues because the environment dynamics are effectively unbounded.

To address this, we apply OOD filtering on the other half of the offline dataset $\mathcal{D}_{\text{GRPO}}$. Specifically, for each observation $o^i_t$, we retain samples where the model-predicted action $\pi^i_{\theta_{\text{SFT}}}(o^i_t)$ either differs from the corresponding dataset action or violates the environment's executable action constraints. Formally, the filtered dataset $\mathcal{D}_{\text{OOD}}$ is defined as:

\begin{equation}
\label{e_5}
\begin{aligned}
\mathcal{D}_{\text{OOD}}
= \Big\{ (o^i_t, a^i_t) \in \mathcal{D}_{\text{GRPO}} \;\Big|\;&
\pi^i_{\theta_{\text{SFT}}}(o^i_t) \neq a^i_t \\
&\text{or}\;\; \pi^i_{\theta_{\text{SFT}}}(o^i_t) \notin \mathcal{A}_{\text{avail}}(o^i_t)
\Big\},
\end{aligned}
\end{equation}

where $\mathcal{A}_{\text{avail}}(o^i_t)$ is the set of available actions based on agent $i$'s local observation $o^i_t$. The filtering procedure is illustrated in Fig.~\ref{f_1}(c). This filtering strategy selectively retains challenging or misaligned samples, allowing subsequent GRPO training to focus on improving robustness against OOD behaviors while reducing overall training cost.

\subsection{Preference Optimization for OOD Generalization}
\label{s_3_4}

To further improve action feasibility and policy alignment, we introduce a preference optimization stage as shown in Fig.~\ref{f_1}(d). Unlike RLHF-style approaches that rely on learning an additional reward model, we adopt simple handcrafted objectives to avoid the challenges of reward estimation in offline multi-agent sequential decision tasks. Specifically, we optimize two criteria: (1) ensuring executability under environment constraints while maintaining generalization, and (2) based on the preferences in the $\mathcal{D}_{\text{OOD}}$ dataset, performance is further improved. The positive preferences correspond to successful trajectories, while the negative preferences correspond to failure trajectories. We define a lightweight preference reward based on these two criteria:

\begin{equation}
\label{e_6}
R(\tau_{\le (o_t^i)}, a_t^i) =
\begin{cases}
\hat{G}_t, & \text{if } a^i_t = \hat{a}^i_t \text{ and } a^i_t \in \mathcal{A}_{\text{avail}}(o^i_t), \\
0, & \text{if } a^i_t \neq \hat{a}^i_t \text{ and } a^i_t \in \mathcal{A}_{\text{avail}}(o^i_t), \\
-1, & \text{if } a^i_t \notin \mathcal{A}_{\text{avail}}(o^i_t),
\end{cases}
\end{equation}

where $\hat{a}^i_t$ denotes the dataset action paired with observation $o_t^i$ in $\mathcal{D}_{\text{OOD}}$. The return-to-go $G_t$ is computed as:

\begin{equation}
G_t = \sum_{k=t}^{T-1} \gamma^{k-t} r_k,
\end{equation}

where $r_k$ is the reward at time step $k$. To ensure stability and prevent large discrepancies in the reward scale, the return-to-go is normalized across the dataset as follows:

\begin{equation}
\hat{G}_t = 2 \cdot \frac{G_t - G_{\min}}{G_{\max} - G_{\min}} - 1,
\end{equation}

where $G_{\min}$ and $G_{\max}$ are the minimum and maximum values of $G_t$ across the dataset, respectively. Given $\mathcal{D}_{\text{OOD}}$ and the corresponding preference rewards, we perform GRPO on the LoRA-adapted DLM-SFT model to refine its decision-making alignment. For each $o^i_t$, we sample a group of $G$ candidate actions $\{a_{j,t}^i\}_{j=1}^G$ from the SFT policy $\pi_{\theta_{\text{SFT}}}(\cdot \mid \tau_{\le(o^i_t)})$. The GRPO loss is formulated as:

\begin{align}
\label{e_7}
\mathcal{L}_{\text{GRPO}} =\ &
\mathbb{E}_{o^i_t \sim \mathcal{D}_{\text{OOD}},\
\{a_{j,t}^i\}_{j=1}^G \sim \pi_{\theta_{\text{SFT}}}(\cdot \mid \tau_{\le(o^i_t)})}
\Bigg\{
\frac{1}{G} \sum_{j=1}^G \frac{1}{|a_{j,t}^i|} \notag \\
& \min\Bigg[
\frac{\pi_{\theta}(a_{j,t}^i \mid \tau_{\le(o^i_t)})}
{\pi_{\theta_{\text{SFT}}}(a_{j,t}^i \mid \tau_{\le(o^i_t)})}
\, \hat{A}_{j,t},
\notag \\
& \operatorname{clip}\!\left(
\frac{\pi_{\theta}(a_{j,t}^i \mid \tau_{\le(o^i_t)})}
{\pi_{\theta_{\text{SFT}}}(a_{j,t}^i \mid \tau_{\le(o^i_t)})},
\, 1-\epsilon,\, 1+\epsilon
\right)\hat{A}_{j,t}
\Bigg] \notag \\
& - \beta\, \mathbb{D}_{KL}\!\left[\pi_{\theta} \,\|\, \pi_{\text{ref}}\right]
\Bigg\},
\end{align}

where $\hat{A}_{j,t}$ denotes the normalized advantage computed from the relative rewards within each sampled group, $|a_{j,t}^i|$ is the sequence length of the $j$-th sampled action, $\epsilon$ is the clipping threshold to ensure training stability, and $\beta$ controls the strength of KL divergence regularization. Through GRPO training, the DLM model achieves improved robustness against OOD actions while maintaining consistency with in-distribution behaviors established during SFT, as detailed in the specific training procedure outlined in Alg.~\ref{a_2}.

\begin{algorithm}[tb]
\caption{GRPO Training Procedure for DLM}
\label{a_2}
\begin{algorithmic}[1]
\STATE \textbf{Input:} Filtered offline dataset $\mathcal{D}_{\text{OOD}}$, fine-tuned parameters $\theta_{\text{SFT}}$, pre-trained LLM $\pi_{\text{ref}}$, max length $L$, number of candidate action groups $G$, clipping threshold $\epsilon$
\STATE \textbf{Initialize:} $\theta_{\text{GRPO}} \leftarrow \theta_{\text{SFT}}$
\FOR{each trajectory $\tau^{(k)} \in \mathcal{D}_{\text{OOD}}$}
  \STATE Tokenize $\tau^{(k)}$ by stacking $(o_t^{(k),i}, a_t^{(k),i})$ with system prompts, truncated to $L$
  \FOR{each agent $i = 1, \dots, N$}
    \STATE Sample $G$ candidate actions $\{a_{j,t}^i\}_{j=1}^G$ from $\pi_{\theta_{\text{SFT}}}(\cdot \mid \tau_{\le (o_t^i)})$
    \STATE Compute advantage $\hat{A}_{j,t}$ for each candidate action $a_j^i$ based on Eq.~\eqref{e_6}
  \ENDFOR
  \STATE Compute the GRPO loss according to Eq.~\eqref{e_7} using the advantage $\hat{A}_{j,t}$
  \STATE Update $\theta_{\text{GRPO}}$ by minimizing the GRPO loss
\ENDFOR
\STATE \textbf{Output:} Fine-tuned model $\theta_{\text{GRPO}}$
\end{algorithmic}
\end{algorithm}

\begin{figure*}
  \centering
  \centerline{\includegraphics[width=0.85\textwidth]{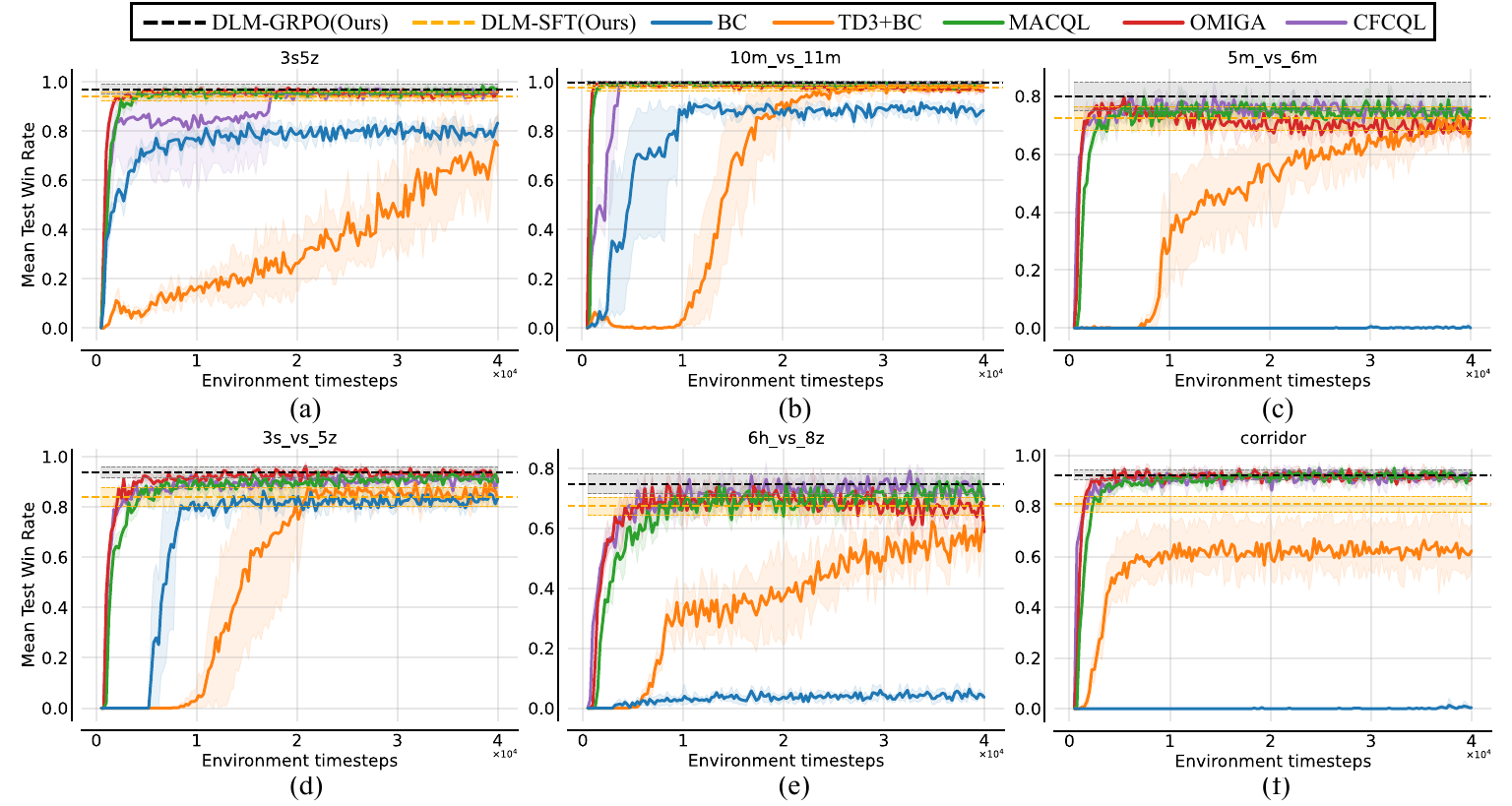}}
  \caption{Performance comparison with value-based offline MARL methods on representative SMAC tasks: (a)-(b) easy, (c)-(d) hard, and (e)-(f) super hard. Only the final test performance of DLM is reported.}
  \label{f_5}
\end{figure*}

\section{Experiment}
\label{s_4}

In this section, we evaluate DLM against a range of offline multi-agent sequential decision-making baselines. We focus comparisons on two categories: (1) value-based offline MARL methods that mitigate distributional shift through value pessimism or action regularization, and (2) LLM-based methods trained via supervised objectives. Specifically, we compare against CFCQL~\citep{shao2024counterfactual}, OMIGA~\citep{wang2024offline}, MACQL~\citep{formanek2024dispelling}, and TD3+BC~\citep{kostrikov2021offline} as value-based baselines, and BC~\citep{syed2008apprenticeship} as an ablation experiment. We also benchmark DLM against LLM-based methods such as MADT~\citep{meng2023offline}, Gato~\citep{reed2022generalist} and SayCan~\citep{ahn2022can} as sequence modeling baselines. Our evaluation spans all tasks in SMAC, a subset of tasks in SMACv2 and LBF, all of which require decentralized coordination under partial observability. We report overall decision quality, analyze improvements in OOD robustness after preference optimization, and assess DLM's zero-shot generalization to unseen tasks. Full experimental setups, including hyperparameter selection and analysis, computational resources, and other implementation details, are provided in Appendix~\ref{s_A_5}, while additional experiments are presented in Appendix~\ref{s_A_6}.

\subsection{Overall Performance on SMAC Benchmark}
\label{s_4_1}

We evaluate DLM across 15 SMAC tasks. The evaluation focuses on mean test win rates as a measure of overall decision quality under decentralized partial observability. All baselines are trained on offline datasets collected following the procedure described in Sec.~\ref{s_3_1}. For fair evaluation, all experiments are conducted with five random seeds, and results are reported as means with a 95\% confidence interval. Note that DLM-SFT and DLM-GRPO use a single model across all tasks, while other baselines are individually trained for each task.
\begin{table*}[t]
  \caption{Performance comparison of LLM-based methods across tasks.}
  \label{t_1}
  \begin{center}
    \begin{small}
      \begin{sc}
        \begin{tabular}{lcccccc}
          \toprule
          \textbf{Method} & 3s5z & 10m\_vs\_11m & 5m\_vs\_6m & 3s\_vs\_5z & 6h\_vs\_8z & corridor\\
          \midrule
          \textbf{MADT} & 0.81$\pm$0.04 & 0.88$\pm$0.03 & 0.67$\pm$0.05 & 0.72$\pm$0.04 & 0.53$\pm$0.05 & 0.64$\pm$0.06 \\
          \textbf{Gato} & 0.72$\pm$0.05 & 0.92$\pm$0.05 & 0.63$\pm$0.04 & 0.65$\pm$0.06 & 0.37$\pm$0.04 & 0.56$\pm$0.05 \\
          \textbf{SayCan} & 0.05$\pm$0.03 & 0.10$\pm$0.05 & 0.07$\pm$0.04 & 0.08$\pm$0.05 & 0.02$\pm$0.01 & 0.00$\pm$0.00 \\
          \textbf{DLM-SFT} & 0.94$\pm$0.02 & 0.98$\pm$0.01 & 0.72$\pm$0.04 & 0.84$\pm$0.04 & 0.67$\pm$0.03 & 0.81$\pm$0.03 \\
          \rowcolor[gray]{0.9} \textbf{DLM-GRPO} & 0.97$\pm$0.02 & 1.00$\pm$0.01 & 0.80$\pm$0.05 & 0.94$\pm$0.02 & 0.75$\pm$0.03 & 0.92$\pm$0.02 \\
          \bottomrule
        \end{tabular}
      \end{sc}
    \end{small}
  \end{center}
  \vskip -0.1in
\end{table*}

\begin{figure}
  \centering
  \centerline{\includegraphics[width=0.85\columnwidth]{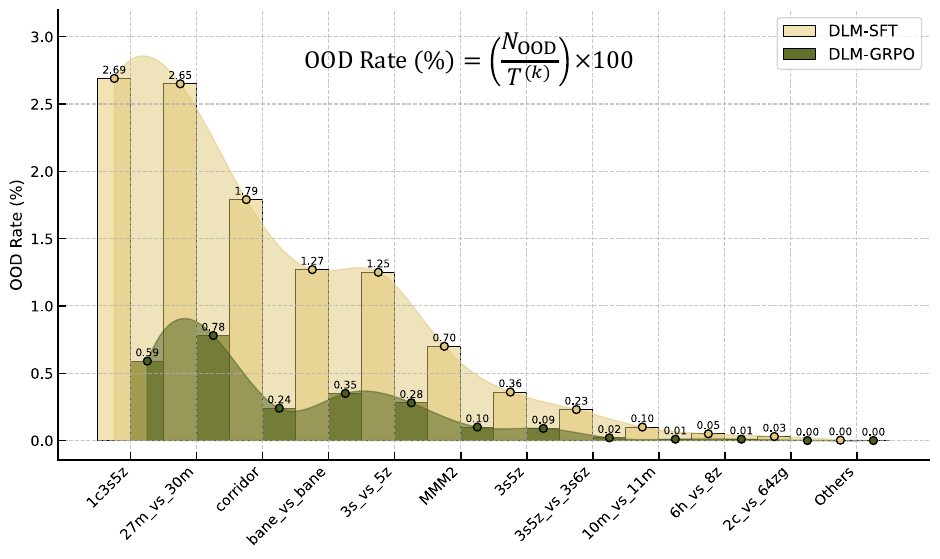}}
  \caption{Comparison of OOD Rates on all SMAC tasks.}
  \label{f_6}
\end{figure}

For value-based offline MARL methods, representative results are shown in Fig.~\ref{f_5}, with full results in Appendix Fig.~\ref{f_A_3}. DLM-GRPO achieves the best average win rates with a single model, matching or surpassing strong baselines such as OMIGA and CFCQL, while DLM-SFT performs comparably to MACQL but remains slightly inferior to DLM-GRPO. TD3+BC and BC perform reasonably on easier tasks but degrade significantly on harder ones, whereas DLM consistently outperforms BC across most settings. This advantage stems from DLM’s dialogue-style sequence construction and multi-agent trajectory alignment, which better preserve inter-agent and temporal decision dependencies. Notably, DLM-SFT, trained without reward supervision, already matches MACQL, and the further gains from DLM-GRPO highlight the effectiveness of preference optimization in improving robustness and decision quality.

For LLM-based methods, as shown in Tab.~\ref{t_1}, SayCan, a zero-shot approach, is evaluated using the LLaMA-3.2-1B-Instruct model with identical parameters for fairness. Across all tasks, DLM variants consistently outperform other LLM-based baselines, with DLM-GRPO achieving the highest win rates. MADT underperforms DLM due to its lack of dialogue-style inter-agent context and the absence of pre-trained LLM priors. Gato shows moderate performance but degrades on more challenging tasks, likely due to the lack of CTDE-aligned dialogue modeling and preference optimization. SayCan performs poorly overall, as the limited capacity of the 1B model hinders effective long-horizon multi-agent decision modeling.

\subsection{OOD Robustness Improvement via Preference Optimization}
\label{s_4_2}

To evaluate the impact of preference optimization, we compare OOD action rates before and after applying GRPO, where OOD actions are defined as those violating environment constraints or deviating from offline behaviors (Sec.\ref{s_3_3}). As shown in Fig.~\ref{f_6}, DLM-SFT, despite its competitive decision quality, occasionally produces OOD actions due to limited data coverage and the absence of action masking. GRPO substantially reduces OOD rates across all tasks, enhancing robustness and execution stability. This reduction is especially pronounced in hard and super hard tasks, where a single OOD action can expose the agent to unseen states and trigger error cascades. A comparison between Fig.~\ref{f_6} and Appendix Fig.~\ref{f_4} reveals that high OOD rates correlate with sparse offline distributions (e.g., \texttt{1c3s5z}, \texttt{27m\_vs\_30m}, \texttt{corridor}), and these tasks also show degraded performance. In contrast, tasks with lower OOD occurrence typically yield higher win rates. These results confirm that preference optimization effectively mitigates OOD risks under limited data and improves generalization without environment interaction.

\subsection{Zero-Shot Generalization to Unseen Tasks}
\label{s_4_3}

\begin{table}
\caption{Zero-shot performance comparison of MADT, DLM-SFT, and DLM-GRPO on unseen tasks.
$^\ast$ Communication tasks in SMAC~\citep{wang2019learning};
$^\dagger$ SMACv2 tasks.}
\label{t_2}
\begin{center}
\begin{small}
\begin{sc}
\setlength{\tabcolsep}{2pt}
\renewcommand{\arraystretch}{0.95}
\begin{tabular}{lccc>{\columncolor[gray]{0.9}}c}
\toprule
\textbf{Task} &  \textbf{MADT} & \textbf{Gato} & \textbf{DLM-SFT} & \textbf{DLM-GRPO} \\
\midrule
3s\_vs\_3z & $0.45$ & $0.52$ & $0.71$ & $0.78$\\
3s\_vs\_4z & $0.73$ & $0.67$ & $0.79$ & $0.82$\\
3m & $0.92$ & $0.88$ & $0.90$ & $0.93$\\
8m & $0.69$ & $0.54$ & $1.00$ & $1.00$\\
25m & $0.57$ & $0.79$ & $0.98$ & $0.99$\\
MMM & $0.85$ & $0.68$ & $0.99$ & $1.00$\\
1o\_10b\_vs\_1r$^\ast$ & $0.13$ & $0.07$ & $0.57$ & $0.64$\\
1o\_2r\_vs\_4r$^\ast$ & $0.11$ & $0.02$ & $0.64$ & $0.69$\\
protoss\_5\_vs\_5$^\dagger$ & $0.00$ & $0.00$ & $0.59$ & $0.67$\\
terran\_5\_vs\_5$^\dagger$ & $0.00$ & $0.00$ & $0.64$ & $0.79$\\
zerg\_5\_vs\_5$^\dagger$ & $0.00$ & $0.00$ & $0.42$ & $0.58$\\
\bottomrule
\end{tabular}
\end{sc}
\end{small}
\end{center}
\vskip -0.1in
\end{table}

We evaluate zero-shot performance on unseen tasks from both SMAC and SMACv2. These tasks are excluded from training and introduce novel unit types, asymmetric team compositions, and coordination patterns. As reported in Tab.~\ref{t_2}, the unified DLM-GRPO model achieves strong win rates on tasks such as \texttt{MMM}, while maintaining competitive performance on more communication-intensive tasks. Although performance decreases in the most challenging SMACv2 tasks, the variation aligns with task similarity to the training distribution: tasks closer to the training data yield higher generalization. Compared with MADT, both DLM-SFT and DLM-GRPO achieve better win rates across nearly all tasks, highlighting the advantage of dialogue-style sequence modeling. In contrast, Gato shows reasonable performance on some simpler SMAC tasks but degrades substantially on communication-heavy scenarios and all SMACv2 tasks, indicating limited transferability to structurally more complex multi-agent settings. Overall, these results confirm that DLM can transfer decision behaviors to structurally novel environments, demonstrating scalability and robustness in zero-shot multi-agent settings.

\subsection{Additional Experimental Analyses}

We conduct additional experiments to validate the effectiveness, robustness, and scalability of DLM, including ablation studies on GRPO and executability-based OOD filtering, analysis of training dynamics and computational cost compared to offline MARL baselines, and evaluation on the LBF benchmark with both in-distribution training results and zero-shot transfer performance on unseen tasks. The specific results are provided in the Appendix.

\section{Conclusion}
\label{s_5}

In this paper, we present DLM, a scalable decision language model for offline multi-agent sequential decision-making across tasks. By reformulating decision processes as dialogue-style sequence modeling, DLM bridges the gap between LLMs and decentralized decision problems. The two-stage training framework, consisting of SFT followed by GRPO, enables DLM to align with environment constraints, mitigate OOD errors, and generalize across tasks with simple handcrafted reward functions. Experiments across multiple benchmarks show that DLM-SFT, trained solely on observations and actions, performs competitively with strong offline baselines. Building on this, DLM-GRPO further enhances robustness and decision quality, outperforming LLM-based methods with a single unified model. Detailed analysis reveals that GRPO effectively reduces OOD action rates, particularly in complex tasks with limited data coverage. DLM also exhibits strong zero-shot generalization to unseen tasks, demonstrating scalability and adaptability. Overall, DLM provides a scalable solution to the generalization bottlenecks in offline MARL.

\section*{Impact Statement}

This work advances offline MARL by improving generalization and robustness across tasks. While DLM is evaluated in simulated environments, it has potential applications in real-world systems, such as autonomous robotics and collaborative decision-making. Future deployment would require attention to computational efficiency and the ability to handle diverse real-world environments. Overall, we do not foresee immediate negative societal impacts.

\bibliography{example_paper}

@article{touvron2023llama,
  title={Llama: Open and efficient foundation language models},
  author={Touvron, Hugo and Lavril, Thibaut and Izacard, Gautier and Martinet, Xavier and Lachaux, Marie-Anne and Lacroix, Timoth{\'e}e and Rozi{\`e}re, Baptiste and Goyal, Naman and Hambro, Eric and Azhar, Faisal and others},
  journal={arXiv preprint arXiv:2302.13971},
  year={2023}
}

@inproceedings{ouyang2022training,
  title={Training language models to follow instructions with human feedback},
  author={Ouyang, Long and Wu, Jeffrey and Jiang, Xu and Almeida, Diogo and Wainwright, Carroll and Mishkin, Pamela and Zhang, Chong and Agarwal, Sandhini and Slama, Katarina and Ray, Alex and others},
  booktitle={Advances in Neural Information Processing Systems},
  volume={35},
  pages={27730--27744},
  year={2022}
}

@article{bai2023qwen,
  title={Qwen technical report},
  author={Bai, Jinze and Bai, Shuai and Chu, Yunfei and Cui, Zeyu and Dang, Kai and Deng, Xiaodong and Fan, Yang and Ge, Wenbin and Han, Yu and Huang, Fei and others},
  journal={arXiv preprint arXiv:2309.16609},
  year={2023}
}

@inproceedings{brown2020language,
  title={Language models are few-shot learners},
  author={Brown, Tom and Mann, Benjamin and Ryder, Nick and Subbiah, Melanie and Kaplan, Jared D and Dhariwal, Prafulla and Neelakantan, Arvind and Shyam, Pranav and Sastry, Girish and Askell, Amanda and others},
  booktitle={Advances in Neural Information Processing Systems},
  volume={33},
  pages={1877--1901},
  year={2020}
}

@article{guo2024deepseek,
  title={DeepSeek-Coder: When the Large Language Model Meets Programming--The Rise of Code Intelligence},
  author={Guo, Daya and Zhu, Qihao and Yang, Dejian and Xie, Zhenda and Dong, Kai and Zhang, Wentao and Chen, Guanting and Bi, Xiao and Wu, Yu and Li, YK and others},
  journal={arXiv preprint arXiv:2401.14196},
  year={2024}
}

@inproceedings{schick2023toolformer,
  title={Toolformer: Language models can teach themselves to use tools},
  author={Schick, Timo and Dwivedi-Yu, Jane and Dess{\`i}, Roberto and Raileanu, Roberta and Lomeli, Maria and Hambro, Eric and Zettlemoyer, Luke and Cancedda, Nicola and Scialom, Thomas},
  booktitle={Advances in Neural Information Processing Systems},
  volume={36},
  pages={68539--68551},
  year={2023}
}

@book{sutton1998reinforcement,
  title={Reinforcement Learning: An Introduction},
  author={Sutton, Richard S and Barto, Andrew G},
  year={1998},
  publisher={MIT Press}
}

@article{schulman2017proximal,
  title={Proximal policy optimization algorithms},
  author={Schulman, John and Wolski, Filip and Dhariwal, Prafulla and Radford, Alec and Klimov, Oleg},
  journal={arXiv preprint arXiv:1707.06347},
  year={2017}
}

@inproceedings{haarnoja2018soft,
  title={Soft Actor-Critic: Off-Policy Maximum Entropy Deep Reinforcement Learning with a Stochastic Actor},
  author={Haarnoja, Tuomas and Zhou, Aurick and Abbeel, Pieter and Levine, Sergey},
  booktitle={Proceedings of the 35th International Conference on Machine Learning},
  pages={1861--1870},
  year={2018},
  volume={80},
}

@inproceedings{kumar2020conservative,
  title={Conservative Q-Learning for Offline Reinforcement Learning},
  author={Kumar, Aviral and Zhou, Aurick and Tucker, George and Levine, Sergey},
  booktitle={Advances in Neural Information Processing Systems},
  volume={33},
  pages={1179--1191},
  year={2020}
}

@article{kostrikov2021offline,
  title={Offline reinforcement learning with implicit q-learning},
  author={Kostrikov, Ilya and Nair, Ashvin and Levine, Sergey},
  journal={arXiv preprint arXiv:2110.06169},
  year={2021}
}

@inproceedings{kumar2019stabilizing,
  title={Stabilizing Off-Policy Q-Learning via Bootstrapping Error Reduction},
  author={Kumar, Aviral and Fu, Justin and Soh, Matthew and Tucker, George and Levine, Sergey},
  booktitle={Advances in Neural Information Processing Systems},
  volume={32},
  pages={4470--4480},
  year={2019}
}

@article{levine2020offline,
  title={Offline reinforcement learning: Tutorial, review, and perspectives on open problems},
  author={Levine, Sergey and Kumar, Aviral and Tucker, George and Fu, Justin},
  journal={arXiv preprint arXiv:2005.01643},
  year={2020}
}

@inproceedings{zhou2024behaviorgpt,
  title={BehaviorGPT: Smart Agent Simulation for Autonomous Driving with Next-Patch Prediction},
  author={Zhou, Zikang and Hu, Haibo and Chen, Xinhong and Wang, Jianping and Guan, Nan and Wu, Kui and Li, Yung-Hui and Huang, Yu-Kai and Xue, Chun Jason},
  booktitle={Advances in Neural Information Processing Systems},
  volume={37},
  pages={79597--79617},
  year={2024}
}

@inproceedings{seraj2023mixed,
  title={Mixed-Initiative Multiagent Apprenticeship Learning for Human Training of Robot Teams},
  author={Seraj, Esmaeil and Xiong, Jerry and Schrum, Mariah and Gombolay, Matthew},
  booktitle={Advances in Neural Information Processing Systems},
  volume={36},
  pages={35426--35440},
  year={2023}
}

@inproceedings{rashid2020weighted,
  title={Weighted QMIX: Expanding Monotonic Value Function Factorisation for Deep Multi-Agent Reinforcement Learning},
  author={Rashid, Tabish and Farquhar, Gregory and Peng, Bei and Whiteson, Shimon},
  booktitle={Advances in Neural Information Processing Systems},
  volume={33},
  pages={10199--10210},
  year={2020}
}

@article{ahn2022can,
  title={Do as i can, not as i say: Grounding language in robotic affordances},
  author={Ahn, Michael and Brohan, Anthony and Brown, Noah and Chebotar, Yevgen and Cortes, Omar and David, Byron and Finn, Chelsea and Fu, Chuyuan and Gopalakrishnan, Keerthana and Hausman, Karol and others},
  journal={arXiv preprint arXiv:2204.01691},
  year={2022}
}

@inproceedings{chen2021decision,
  title={Decision Transformer: Reinforcement Learning via Sequence Modeling},
  author={Chen, Lili and Lu, Kevin and Rajeswaran, Aravind and Lee, Kimin and Grover, Aditya and Laskin, Misha and Abbeel, Pieter and Srinivas, Aravind and Mordatch, Igor},
  booktitle={Advances in Neural Information Processing Systems},
  volume={34},
  pages={15084--15097},
  year={2021}
}

@inproceedings{janner2021offline,
  title={Offline Reinforcement Learning as One Big Sequence Modeling Problem},
  author={Janner, Michael and Li, Qiyang and Levine, Sergey},
  booktitle={Advances in Neural Information Processing Systems},
  volume={34},
  pages={1273--1286},
  year={2021}
}

@article{meng2023offline,
  title={Offline Pre-Trained Multi-Agent Decision Transformer},
  author={Meng, Linghui and Wen, Muning and Le, Chenyang and Li, Xiyun and Xing, Dengpeng and Zhang, Weinan and Wen, Ying and Zhang, Haifeng and Wang, Jun and Yang, Yaodong},
  journal={Machine Intelligence Research},
  volume={20},
  number={2},
  pages={233--248},
  year={2023},
}

@inproceedings{zhang2025bridging,
  title={Bridging training and execution via dynamic directed graph-based communication in cooperative multi-agent systems},
  author={Zhang, Zhuohui and He, Bin and Cheng, Bin and Li, Gang},
  booktitle={Proceedings of the AAAI Conference on Artificial Intelligence},
  volume={39},
  pages={23395--23403},
  year={2025}
}

@inproceedings{fujimoto2019off,
  title={Off-Policy Deep Reinforcement Learning Without Exploration},
  author={Fujimoto, Scott and Meger, David and Precup, Doina},
  booktitle={Proceedings of the 36th International Conference on Machine Learning},
  pages={2052--2062},
  year={2019},
  volume={97},
}

@inproceedings{lowe2017multi,
  title={Multi-Agent Actor-Critic for Mixed Cooperative-Competitive Environments},
  author={Lowe, Ryan and Wu, Yi I and Tamar, Aviv and Harb, Jean and Abbeel, Pieter and Mordatch, Igor},
  booktitle={Advances in Neural Information Processing Systems},
  volume={30},
  pages={6379--6390},
  year={2017}
}

@article{shao2024deepseekmath,
  title={Deepseekmath: Pushing the limits of mathematical reasoning in open language models},
  author={Shao, Zhihong and Wang, Peiyi and Zhu, Qihao and Xu, Runxin and Song, Junxiao and Bi, Xiao and Zhang, Haowei and Zhang, Mingchuan and Li, YK and Wu, Y and others},
  journal={arXiv preprint arXiv:2402.03300},
  year={2024}
}

@article{samvelyan2019starcraft,
  author={Mikayel Samvelyan and Tabish Rashid and Christian Schroeder De Witt and Gregory Farquhar and Nantas Nardelli and Tim GJ Rudner and Chia-Man Hung and Philip HS Torr and Jakob Foerster and Shimon Whiteson},
  title={The Starcraft Multi-Agent Challenge},
  journal={arXiv:1902.04043},
  year={2019}
}

@inproceedings{syed2008apprenticeship,
  title={Apprenticeship Learning Using Linear Programming},
  author={Syed, Umar and Bowling, Michael and Schapire, Robert E},
  booktitle={Proceedings of the 25th International Conference on Machine Learning},
  pages={1032--1039},
  year={2008}
}

@inproceedings{wang2024offline,
  title={Offline Multi-Agent Reinforcement Learning with Implicit Global-to-Local Value Regularization},
  author={Wang, Xiangsen and Xu, Haoran and Zheng, Yinan and Zhan, Xianyuan},
  booktitle={Advances in Neural Information Processing Systems},
  volume={37},
  pages={52413--52429},
  year={2024}
}

@inproceedings{shao2024counterfactual,
  title={Counterfactual Conservative Q Learning for Offline Multi-Agent Reinforcement Learning},
  author={Shao, Jianzhun and Qu, Yun and Chen, Chen and Zhang, Hongchang and Ji, Xiangyang},
  booktitle={Advances in Neural Information Processing Systems},
  volume={36},
  pages={77290--77312},
  year={2023}
}

@inproceedings{formanek2024dispelling,
  title={Dispelling the Mirage of Progress in Offline MARL through Standardised Baselines and Evaluation},
  author={Formanek, Juan and Tilbury, Callum R and Beyers, Louise and Shock, Jonathan and Pretorius, Arnu},
  booktitle={Advances in Neural Information Processing Systems},
  volume={37},
  pages={139650--139672},
  year={2024}
}

@inproceedings{vaswani2017attention,
  title={Attention Is All You Need},
  author={Vaswani, Ashish and Shazeer, Noam and Parmar, Niki and Uszkoreit, Jakob and Jones, Llion and Gomez, Aidan N and Kaiser, {\L}ukasz and Polosukhin, Illia},
  booktitle={Advances in Neural Information Processing Systems},
  volume={30},
  pages={5998--6008},
  year={2017}
}

@article{reed2022generalist,
  title={A generalist agent},
  author={Reed, Scott and Zolna, Konrad and Parisotto, Emilio and Colmenarejo, Sergio Gomez and Novikov, Alexander and Barth-Maron, Gabriel and Gimenez, Mai and Sulsky, Yury and Kay, Jackie and Springenberg, Jost Tobias and others},
  journal={arXiv preprint arXiv:2205.06175},
  year={2022}
}

@article{hu2021lora,
  title={Lora: Low-rank adaptation of large language models},
  author={Hu, Edward J and Shen, Yelong and Wallis, Phillip and Allen-Zhu, Zeyuan and Li, Yuanzhi and Wang, Shean and Wang, Lu and Chen, Weizhu},
  journal={arXiv:2106.09685},
  year={2021}
}

@article{zhang2025pagnet,
  title={PAGNet: Pluggable Adaptive Generative Networks for Information Completion in Multi-Agent Communication},
  author={Zhang, Zhuohui and Cheng, Bin and Wang, Zhipeng and Zhou, Yanmin and Li, Gang and Lu, Ping and He, Bin and Chen, Jie},
  journal={arXiv preprint arXiv:2502.03845},
  year={2025}
}

@article{grattafiori2024llama,
  title={The llama 3 herd of models},
  author={Grattafiori, Aaron and Dubey, Abhimanyu and Jauhri, Abhinav and Pandey, Abhinav and Kadian, Abhishek and Al-Dahle, Ahmad and Letman, Aiesha and Mathur, Akhil and Schelten, Alan and Vaughan, Alex and others},
  journal={arXiv preprint arXiv:2407.21783},
  year={2024}
}

@book{oliehoek2016concise,
  title={A Concise Introduction to Decentralized POMDPs},
  author={Oliehoek, Frans A and Amato, Christopher},
  year={2016},
  publisher={Springer}
}

@inproceedings{mikolov2013distributed,
  title={Distributed Representations of Words and Phrases and Their Compositionality},
  author={Mikolov, Tomas and Sutskever, Ilya and Chen, Kai and Corrado, Greg S and Dean, Jeff},
  booktitle={Advances in Neural Information Processing Systems},
  volume={26},
  pages={3111--3119},
  year={2013}
}

@article{fu2020d4rl,
  author={Justin Fu and Aviral Kumar and Ofir Nachum and George Tucker and Sergey Levine},
  title={{D4RL: Datasets for Deep Data-Driven Reinforcement Learning}},
  journal={arXiv:2004.07219},
  year={2020}
}

@article{mixrts,
  title={MIXRTs: Toward Interpretable Multi-Agent Reinforcement Learning via Mixing Recurrent Soft Decision Trees},
  author={Liu, Zichuan and Zhu, Yuanyang and Wang, Zhi and Gao, Yang and Chen, Chunlin},
  journal={IEEE Transactions on Pattern Analysis and Machine Intelligence},
  volume={47},
  number={5},
  pages={4090--4107},
  year={2025}
}

@article{wang2019learning,
  title={Learning nearly decomposable value functions via communication minimization},
  author={Wang, Tonghan and Wang, Jianhao and Zheng, Chongyi and Zhang, Chongjie},
  journal={arXiv preprint arXiv:1910.05366},
  year={2019}
}

@article{formanek2023off,
  title={Off-the-Grid MARL: Datasets with Baselines for Offline Multi-Agent Reinforcement Learning},
  author={Formanek, Claude and Jeewa, Asad and Shock, Jonathan and Pretorius, Arnu},
  journal={arXiv preprint arXiv:2302.00521},
  year={2023}
}

@book{cover2006elements,
  title={Elements of Information Theory},
  author={Cover, Thomas M and Thomas, Joy A},
  year={2006},
  publisher={Wiley-Interscience},
  edition={2nd}
}

@article{papoudakis2020benchmarking,
  title={Benchmarking Multi-Agent Deep Reinforcement Learning Algorithms in Cooperative Tasks},
  author={Papoudakis, Georgios and Christianos, Filippos and Sch{\"a}fer, Lukas and Albrecht, Stefano V},
  journal={arXiv preprint arXiv:2006.07869},
  year={2020}
}

@inproceedings{ellis2024smacv2,
  title={SMACv2: An Improved Benchmark for Cooperative Multi-Agent Reinforcement Learning},
  author={Ellis, Benjamin and Cook, Jonathan and Moalla, Skander and Samvelyan, Mikayel and Sun, Mingfei and Mahajan, Anuj and Foerster, Jakob and Whiteson, Shimon},
  booktitle={Proc. 37th Conf. Neural Inf. Process. Syst.},
  address={New Orleans, LA, USA},
  month={Dec.},
  year={2023},
  pages={22361--22369}
}

@article{rangwala2020learning,
  title={Learning multi-agent communication through structured attentive reasoning},
  author={Rangwala, Murtaza and Williams, Ryan},
  journal={Advances in Neural Information Processing Systems},
  volume={33},
  pages={10088--10098},
  year={2020}
}
\bibliographystyle{icml2026}

\newpage
\appendix
\onecolumn
\section{Appendix}

\subsection{Supplementary Figures and Algorithms}
\paragraph{Inference Procedure for DLM}
In Sec.~\ref{s_3_2} of the main paper, we describe the inference process for the DLM, which enables decentralized decision-making based on a dialogue-style sequence modeling approach. The specific inference procedure is outlined in Alg.~\ref{a_3}, where each agent independently generates actions based on its local observation history. Unlike prior single-agent formulations, DLM uses a centralized training framework but operates in a decentralized manner during inference, allowing agents to generate actions from their individual perspectives.

\begin{algorithm}
\caption{Inference Procedure for DLM}
\label{a_3}
\begin{algorithmic}[1]
\STATE \textbf{Input:} Fine-tuned model $\theta_{\text{SFT}}$, tokenizer, environment
\FOR{each episode}
  \STATE Reset environment and initialize history
  \WHILE{episode not terminated}
    \STATE Encode observations into prompts
    \FOR{each agent $i=1,\dots,N$}
      \STATE Predict action according to Eq.~\eqref{e_3}
      \IF{action invalid or unavailable}
        \STATE Resample from Eq.~\eqref{e_4}
      \ENDIF
    \ENDFOR
    \STATE Execute actions and update history
  \ENDWHILE
  \STATE Record episode outcome
\ENDFOR
\STATE \textbf{Output:} Test win rates
\end{algorithmic}
\end{algorithm}

\begin{figure}[h]
  \centering
  \centerline{\includegraphics[width=0.45\columnwidth]{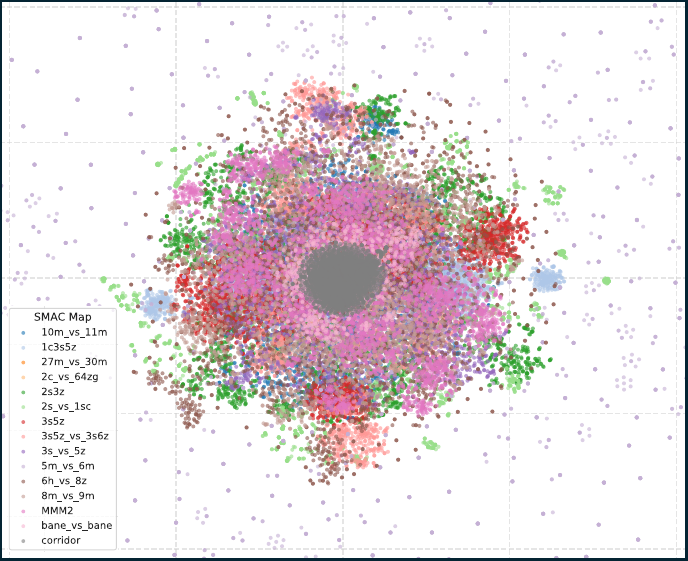}}
  \caption{t-SNE projection of observation distributions from the offline dataset across all SMAC tasks.}
  \label{f_4}
\end{figure}

\paragraph{t-SNE Projection}
In Sec.~\ref{s_3_3} of the main paper, we illustrate the challenge of OOD actions using t-SNE visualization. As shown in Fig.~\ref{f_4}, even with diverse trajectory collection across multiple tasks, the observations and actions sampled from the offline dataset occupy only a sparse subset of the overall space. This emphasizes the inherent limitation of dataset coverage and demonstrates that enlarging the dataset alone cannot fully eliminate OOD issues due to the unbounded nature of the environment's dynamics.

\subsection{Dataset Construction Details}
\label{s_A_1}

We explain the dataset construction process using SMAC as an example. The same procedure is applied to other benchmarks, ensuring consistency across all tasks.

\paragraph{Data Collection} 

To ensure the credibility and reproducibility of our experimental results, we initially explored the use of publicly available offline MARL datasets. While several recent datasets, such as OG-MARL~\citep{formanek2023off}, provide high-quality offline trajectories for MARL, they typically cover only a limited subset of SMAC~\citep{samvelyan2019starcraft} tasks and do not support comprehensive multi-task evaluation. To address this limitation, we construct our own dataset following the data collection methodology of D4RL~\citep{fu2020d4rl}. Specifically, we adopt TGCNet~\citep{zhang2025bridging} as the behavior policy for data collection, due to its ability to achieve near-perfect performance across all SMAC tasks. For each task, we train TGCNet in an online setting and apply early stopping once the win rate exceeds 80\%, using a test interval of 50000 steps. The resulting checkpoint is then used to interact with the environment and collect 4000 high-quality trajectories per task. Each collected trajectory includes standard components commonly used in offline MARL benchmarks: \texttt{actions}, \texttt{actions\_onehot}, \texttt{avail\_actions}, \texttt{filled}, \texttt{obs}, \texttt{reward}, \texttt{state} and \texttt{terminated}. As detailed in Sec.3.1, this process yields a diverse and consistent dataset spanning all 15 SMAC tasks. For DLM, the dataset is split into two subsets, one for SFT and the other for GRPO. Importantly, all baseline algorithms, including both value-based and imitation-based methods, are trained using the full dataset without any modifications. Therefore, although DLM employs a two-stage training procedure with a dataset split, it uses the same total amount of data as other baselines. All methods operate on an identical data distribution to ensure fairness and comparability.

\paragraph{Dataset Quality}

To verify the reliability and quality of the collected trajectories, we conduct a quantitative analysis of the dataset. For each SMAC task, we divide the 4000 collected episodes evenly into two subsets of 2000 trajectories. This partition supports the two-stage training procedure of DLM, where one subset is used for SFT and the other for GRPO. We compute the average return and standard deviation for each subset by summing the per-step rewards within each episode and then aggregating statistics over 2000 trajectories. The results are presented in Tab.~\ref{t_A_1}, which reports the mean $\pm$ standard deviation of episode returns for every task in both subsets, along with the overall average across all tasks. Although the maximum achievable return varies by task depending on episode length and reward sparsity, most of the collected trajectories yield returns consistent with the threshold used in our early-stopping strategy, where trajectory generation begins once TGCNet reaches a win rate of at least 80\%. This confirms that the dataset meets the expected quality standard and is suitable for training reliable offline policies.

\begin{table}[h]
\centering
\caption{Trajectory return (mean $\pm$ std) per SMAC task in the collected offline dataset. Each map contains 2000 trajectories divided equally into two subsets.}
\label{t_A_1}
\begin{tabular}{lccc}
\toprule
\textbf{Task} & \textbf{SFT Subset} & \textbf{GRPO Subset} & \textbf{Total Dataset} \\
\midrule
2s\_vs\_1sc        & 19.18 $\pm$ 2.78 & 19.17 $\pm$ 2.79 & 19.17 $\pm$ 2.78 \\
2s3z              & 19.89 $\pm$ 0.89 & 19.89 $\pm$ 0.88 & 19.89 $\pm$ 0.89 \\
3s5z              & 19.74 $\pm$ 1.19 & 19.76 $\pm$ 1.08 & 19.75 $\pm$ 1.14 \\
1c3s5z            & 19.98 $\pm$ 0.44 & 19.97 $\pm$ 0.55 & 19.97 $\pm$ 0.50 \\
10m\_vs\_11m      & 19.79 $\pm$ 1.15 & 19.72 $\pm$ 1.35 & 19.75 $\pm$ 1.25 \\
2c\_vs\_64zg      & 20.00 $\pm$ 1.12 & 19.99 $\pm$ 1.14 & 19.99 $\pm$ 1.12 \\
5m\_vs\_6m        & 18.28 $\pm$ 3.74 & 18.35 $\pm$ 3.59 & 18.34 $\pm$ 3.66 \\
bane\_vs\_bane    & 20.00 $\pm$ 0.00 & 20.00 $\pm$ 0.00 & 20.00 $\pm$ 0.00 \\
3s\_vs\_5z        & 21.64 $\pm$ 1.33 & 21.62 $\pm$ 1.39 & 21.63 $\pm$ 1.36 \\
8m\_vs\_9m        & 19.07 $\pm$ 2.57 & 19.04 $\pm$ 2.58 & 19.06 $\pm$ 2.58 \\
3s5z\_vs\_3s6z    & 19.83 $\pm$ 1.14 & 19.84 $\pm$ 1.07 & 19.83 $\pm$ 1.11 \\
27m\_vs\_30m      & 19.17 $\pm$ 2.20 & 19.25 $\pm$ 2.20 & 19.21 $\pm$ 2.20 \\
6h\_vs\_8z        & 18.46 $\pm$ 2.47 & 18.45 $\pm$ 2.50 & 18.45 $\pm$ 2.48 \\
MMM2              & 19.41 $\pm$ 1.96 & 19.38 $\pm$ 1.99 & 19.40 $\pm$ 1.98 \\
corridor          & 19.76 $\pm$ 2.11 & 19.69 $\pm$ 2.22 & 19.72 $\pm$ 2.17 \\
\midrule
\textbf{Average}  & \textbf{19.61 $\pm$ 1.67} & \textbf{19.60 $\pm$ 1.68} & \textbf{19.61 $\pm$ 1.68} \\
\bottomrule
\end{tabular}
\end{table}

\paragraph{Dialogue-Style Conversion}

\begin{figure}
  \centering
  \centerline{\includegraphics[width=0.8\columnwidth]{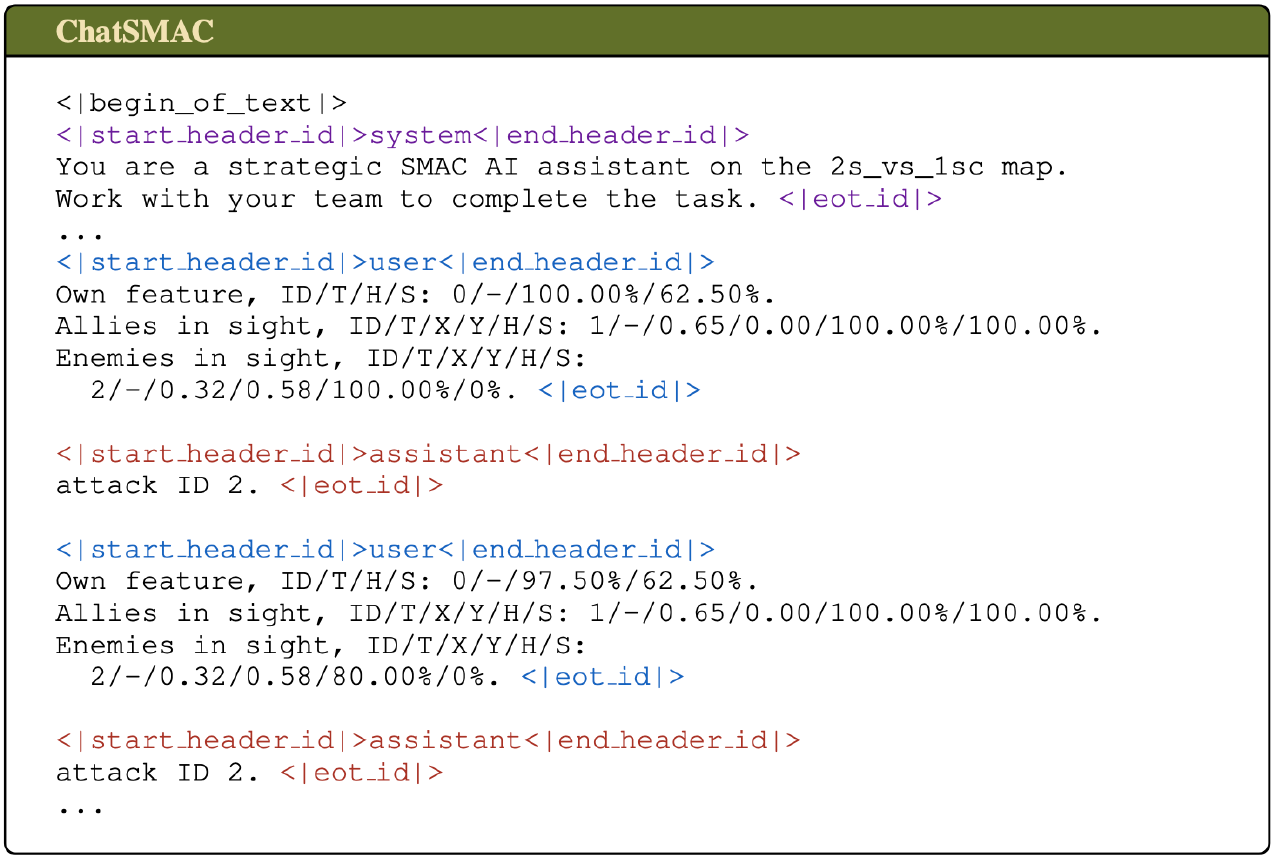}}
  \caption{An example trajectory from the ChatSMAC dataset formatted in chat format.}
  \label{f_A}
\end{figure}

For training DLM, we apply an additional transformation to the original offline dataset, converting it into a dialogue-style format suitable for multi-turn sequence modeling. In this process, we retain only the \texttt{obs} and \texttt{actions} fields at each timestep and organize them into observation–action pairs. Each pair is then verbalized as a natural language dialogue turn, as illustrated in Fig.~\ref{f_2} of the main text. We begin by constructing the \textit{system prompt}, which corresponds to the system role in the chat interface and provides high-level scenario context. Since each SMAC task corresponds to a specific map, we generate map-specific instructions in the form: \texttt{"You are a strategic SMAC AI assistant on the \_\_ map. Work with your team to complete the task."} This prompt guides the model to behave as a cooperative agent grounded in the given environment. Next, we generate the \textit{observation prompt}, which corresponds to the user role in the chat interaction and encodes the agent’s local observation at the current timestep. To construct this prompt, we assign each agent a unique ID and extract key features from its observation vector. The agent’s own attributes, such as ID, unit type, health, and shield, are obtained from the \texttt{own\_feats} field. Information about allied units within the agent’s sight is retrieved from \texttt{ally\_feats}, which includes their IDs, types, relative positions in X and Y coordinates, health, and shield. Likewise, the \texttt{enemy\_feats} field provides corresponding information for visible enemies. These features are verbalized into a structured natural language description that captures the agent’s local perspective. Finally, we construct the \textit{action prompt}, which corresponds to the assistant role in the chat interaction and represents the agent’s response based on its selected action. Since actions in the dataset are represented as discrete indices, we first decode each index based on the SMAC action mapping. For example, action 0 corresponds to \texttt{"no-op"}, action 1 to \texttt{"stop"}, and action 2 to \texttt{"move north one step"}. The decoded action is then used as the assistant's response, completing the observation–action dialogue turn. Each complete turn is wrapped using the chat format adopted by LLaMA-3~\citep{grattafiori2024llama}, ensuring compatibility with mainstream language models. This process produces the final dialogue-style dataset, which we refer to as ChatSMAC. It serves as the foundation for training DLM. Full examples of prompt construction and template specifications are provided in Fig.~\ref{f_A}.

\subsection{Details about Benchmarks}
\label{s_A_2}

\paragraph{SMAC Overview}

The SMAC~\citep{samvelyan2019starcraft} benchmark, built upon the StarCraft II engine, has been widely used for evaluating cooperative multi-agent reinforcement learning. It focuses on micromanagement tasks where each agent controls a single unit and must coordinate with others under partial observability and sparse global rewards. Recently, SMACv2~\citep{ellis2024smacv2} was introduced to address the limitations of SMAC by introducing additional randomness, restricting agents’ field of view, and providing more diverse and challenging tasks, making it a more rigorous testbed for assessing generalization, robustness, and scalability of multi-agent learning methods.

\paragraph{Task Grouping}

\begin{figure}
  \centering
  \centerline{\includegraphics[width=\textwidth]{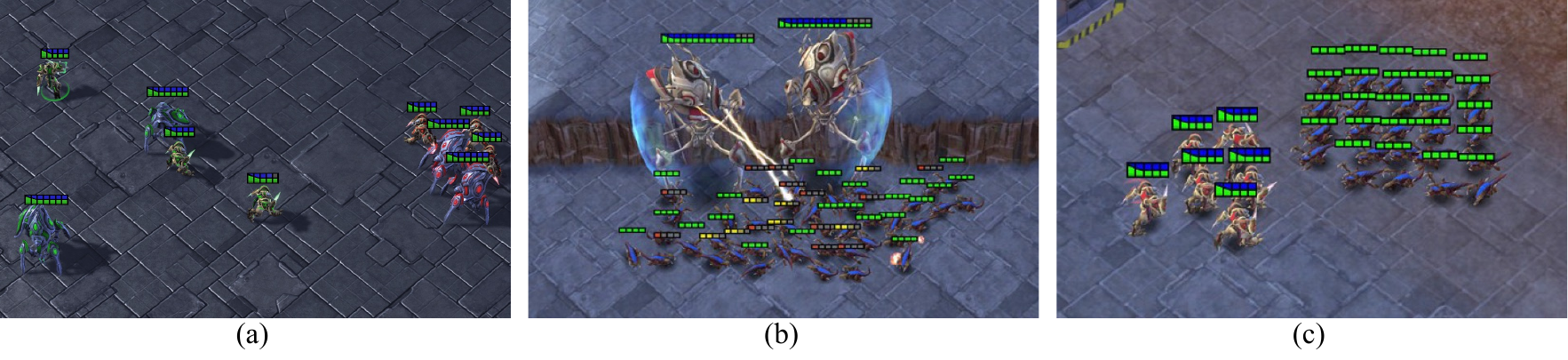}}
  \caption{Screenshots of SMAC tasks at different difficulty levels: (a) \texttt{2s3z} (easy), (b) \texttt{2c\_vs\_64zg} (hard), and (c) \texttt{corridor} (super hard).}
  \label{f_A_1}
\end{figure}

SMAC tasks are commonly divided into three difficulty levels: easy, hard, and super hard. This classification is based on factors such as unit types, asymmetry between teams, and the complexity of required coordination. Easy tasks typically involve symmetric unit compositions and can often be solved with basic strategies. In contrast, hard and super hard tasks introduce heterogeneous units, asymmetric team settings, and demand more sophisticated tactics such as precise positioning, focus firing, and kiting. Representative examples of each difficulty level are illustrated in Fig.~\ref{f_A_1}, highlighting the increasing complexity across categories. A full list of task groupings by difficulty is summarized in Tab.~\ref{t_A_2}.

\begin{table}
\centering
\caption{SMAC task difficulty classification.}
\label{t_A_2}
\begin{tabular}{ll}
\toprule
\textbf{Difficulty Level} & \textbf{Task} \\
\midrule
\multirow{5}{*}{Easy} 
& 2s\_vs\_1sc \\
& 2s3z \\
& 3s5z \\
& 1c3s5z \\
& 10m\_vs\_11m \\
\midrule
\multirow{5}{*}{Hard} 
& 2c\_vs\_64zg \\
& 5m\_vs\_6m \\
& bane\_vs\_bane \\
& 3s\_vs\_5z \\
& 8m\_vs\_9m \\
\midrule
\multirow{5}{*}{Super Hard} 
& 3s5z\_vs\_3s6z \\
& 27m\_vs\_30m \\
& 6h\_vs\_8z \\
& MMM2 \\
& corridor \\
\bottomrule
\end{tabular}
\end{table}

\paragraph{Observation and Action Spaces}

In each SMAC task, agents receive low-dimensional local observations that encode information about nearby allies, enemies, and the agent’s own state. The observation space has a fixed dimensionality, but the content depends on the number of visible units, reflecting the partial observability of the environment. The action space is discrete and includes primitive operations such as moving in four directions, attacking or healing specific enemies or allies, stopping, and executing a no-operation. Notably, the set of available actions is dynamic, determined by the agent’s current local visibility and unit-specific constraints.



\paragraph{Choice of Multi-Task Benchmarks}

We adopt SMAC~\citep{samvelyan2019starcraft} and its extension SMACv2~\citep{ellis2024smacv2} as the primary benchmarks for evaluating DLM in a multi-task setting. Following prior work such as MADT~\citep{meng2023offline}, each task within SMAC is conventionally treated as a distinct task, since maps differ in agent types and numbers, available abilities, team compositions, and coordination requirements. This diversity spans simple symmetric battles to highly asymmetric matchups that demand fine-grained cooperation, thereby aligning with common definitions of multi-task reinforcement learning as learning across a distribution of environments with varied state/action spaces and task goals. Compared with SMAC, SMACv2 introduces additional randomness, restricted fields of view, and more heterogeneous unit compositions, which further increase task variability and difficulty. Together, SMAC and SMACv2 provide a scalable and reproducible platform where heterogeneous cooperative tasks can be systematically evaluated under a unified framework, making them suitable for assessing the generalization capacity of large decision models like DLM.

\subsection{Motivation Behind the Design}
\label{app:design_motivation}

A central component of DLM is the representation of multi-agent trajectories as sequences of observation--action pairs:
\begin{equation}
\label{e_1_recap}
\tau^{(k)} = \left( \, \left\{ (o_t^{(k),i},\ a_t^{(k),i})\ \big|\ t = 1,\ldots, T^{(k)} \right\} \, \right)_{i=1}^{N},
\end{equation}
where $o_t^{(k),i}$ and $a_t^{(k),i}$ denote the local observation and executed action of agent $i$ at timestep $t$ in the $k$-th trajectory, and $T^{(k)}$ is the episode length. This design is motivated by a careful analysis of the limitations of BC in multi-agent settings.

\paragraph{Limitation 1: Absence of Inter-Agent Information}

Conventional BC learns a local policy $\pi^i(o^i) = p(a \mid o^i)$ by mapping the agent’s private observation $o^i$ to an action $a$, while ignoring critical dependencies on other agents' information. In cooperative multi-agent tasks, a common idealization is to treat all agents as components of a single joint agent operating over the full global state $s$, in which case the optimal policy is defined as $\pi^i(s) = p(a \mid s)$. However, under partial observability, the agent's local observation $o^i$ may correspond to multiple possible global states, causing $p(a \mid o^i)$ to become a weighted mixture over the optimal policies for different $s$. This discrepancy leads to a mismatch between $p(a \mid o^i)$ and the true optimal policy $p(a \mid s)$. Formally, we can express this mismatch as:
\begin{equation}
\label{eq:representation_gap_1}
p(a \mid o^i) = \sum_s p(a \mid s) \cdot p(s \mid o^i),
\end{equation}
where the posterior $p(s \mid o^i)$ represents a distribution over global states consistent with $o^i$. When $p(s \mid o^i)$ has high entropy, the resulting policy becomes a blurred mixture, leading to suboptimal actions. Now consider conditioning on the joint observations of all agents $(o^1, \dots, o^n)$. The corresponding policy is:
\begin{equation}
\label{eq:representation_gap_2}
p(a \mid o^1, \dots, o^n) = \sum_s p(a \mid s) \cdot p(s \mid o^1, \dots, o^n).
\end{equation}
From the information-theoretic property that conditioning reduces entropy~\citep{cover2006elements}, we have:
\begin{equation}
\mathcal{H}(s \mid o^i) \geq \mathcal{H}(s \mid o^1, \dots, o^n),
\end{equation}
which implies that the posterior $p(s \mid o^1, \dots, o^n)$ is more concentrated than $p(s \mid o^i)$.

As a result, the weighted average in Eq.~\ref{eq:representation_gap_2} more closely approximates the true optimal policy $p(a \mid s)$ than Eq.~\ref{eq:representation_gap_1}. This analysis highlights that incorporating full agent trajectories, as in our dialogue-style formulation, reduces the representation gap and leads to higher-quality decision making compared to conventional BC.

\paragraph{Limitation 2: Lack of Temporal Dependency}

BC also suffers from ignoring temporal dependencies by treating each timestep independently, i.e., modeling the policy as $p(a_t^i \mid o_t^i)$ without incorporating past observations or actions. However, in partially observable environments, the current observation $o_t^i$ alone is generally insufficient to infer the true latent state of the environment. As a result, this memoryless policy lacks the contextual information required for strategic reasoning over time.

To formalize this limitation, consider that the optimal policy in a Dec-POMDP depends on the full action-observation history $h_t^i = (o_1^i, a_1^i, \dots, o_{t-1}^i, a_{t-1}^i, o_t^i)$. The true optimal policy is therefore:
\begin{equation}
    \pi^i_{\text{opt}} = p(a_t^i \mid h_t^i),
\end{equation}
whereas BC approximates this as:
\begin{equation}
    \pi^i_{\text{BC}} = p(a_t^i \mid o_t^i).
\end{equation}
Applying the data processing inequality~\citep{cover2006elements}, we know that:
\begin{equation}
    I(a_t^i; h_t^i) \geq I(a_t^i; o_t^i),
\end{equation}
where $I(\cdot\,;\cdot)$ denotes mutual information. This inequality highlights that conditioning on full history provides strictly more information about the optimal action than conditioning on $o_t^i$ alone.

Thus, ignoring historical context reduces the model's capacity to learn strategies that rely on long-term planning, multi-agent coordination, or temporal disambiguation. Such strategies are frequently required in complex tasks, including kiting, flanking maneuvers, or delayed action execution. DLM addresses this limitation by formulating decision-making as an autoregressive sequence modeling problem. By retaining the full sequence of past observation–action pairs as dialogue history, the model can effectively leverage long-range temporal dependencies to make more informed and coherent decisions.

\paragraph{Design Motivation of DLM}
To address these limitations, DLM reformulates the decision process as a dialogue-style sequence modeling problem. Instead of fitting per-timestep policies independently, it models the entire trajectory as a structured autoregressive sequence of $(o, a)$ pairs. This design allows:
\begin{itemize}
  \item \textbf{Contextual Encoding}: Inter-agent relationships are captured implicitly in the structured dialogue, where each agent's input includes both its own and nearby agents' attributes, encoded in natural language.
  \item \textbf{Temporal Dependency}: By modeling decisions autoregressively, the model naturally learns from the accumulated context of previous observations and actions, thus capturing history without explicit recurrence.
  \item \textbf{CTDE Compatibility}: Representing trajectories in language format naturally supports the CTDE paradigm. Each dialogue-style trajectory encodes the full decision process of an individual agent, while maintaining access to global information across agents during training. This enables the model to learn coordinated strategies centrally, yet make decisions based solely on local observations during execution.

\end{itemize}

In summary, the design of Eq.~\ref{e_1_recap} serves as a unified interface that preserves agent-level autonomy while enabling temporally and contextually grounded decision-making. This structure is essential for bridging the gap between language models and multi-agent sequential decision processes.

\subsection{Training and Implementation Details}
\label{s_A_5}

\begin{figure}
  \centering
  \centerline{\includegraphics[width=0.9\textwidth]{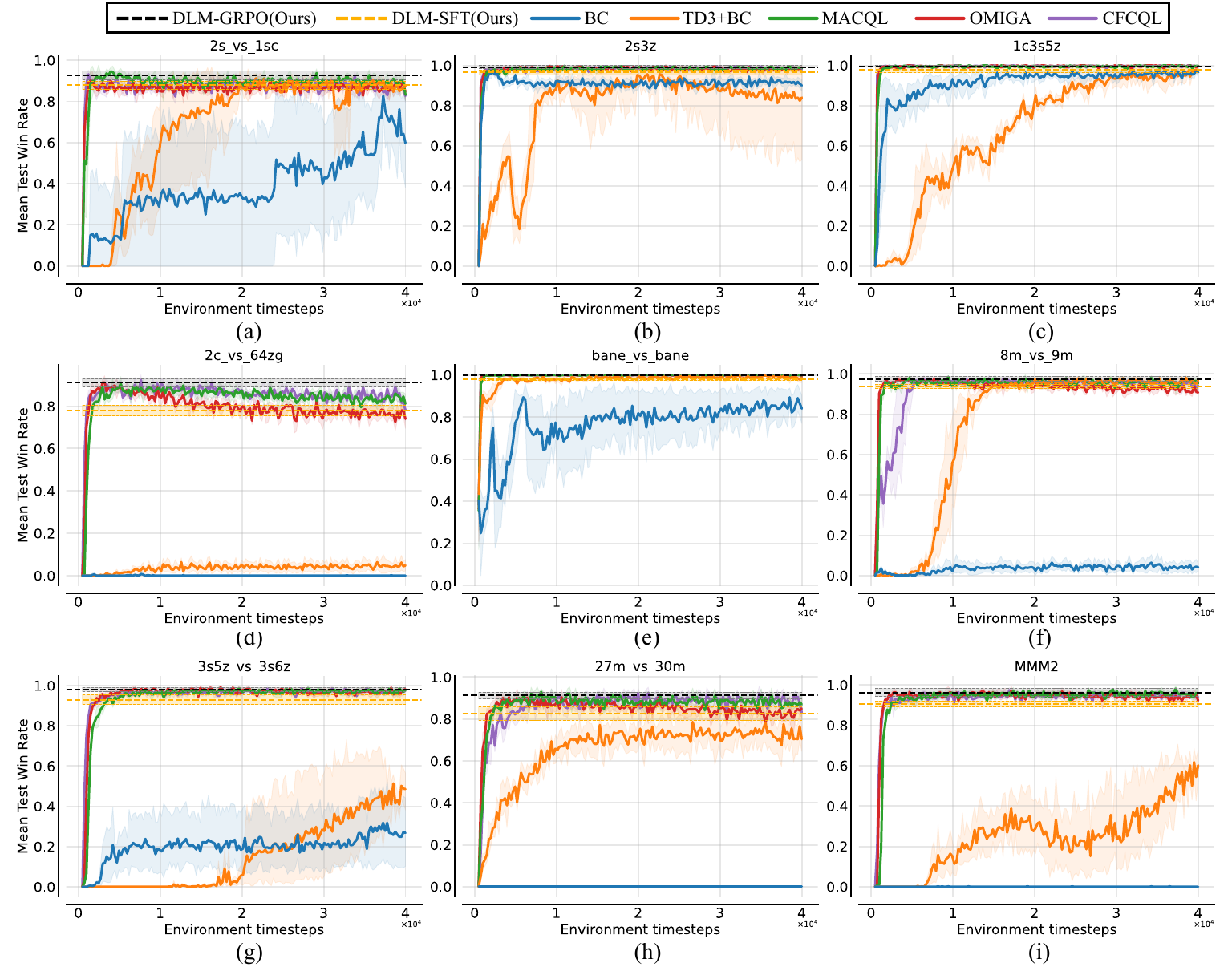}}
  \caption{Performance comparison with baselines on remaining SMAC tasks: (a)-(c) easy, (d)-(f) hard, and (g)-(i) super hard. Only the final test performance of DLM is reported.}
  \label{f_A_3}
\end{figure}

\paragraph{Supplementary Results for Sec.~\ref{s_4_1}}
\label{s_A_5_1}
In Sec.~\ref{s_4_1}, we reported results on six representative tasks from the SMAC benchmark. Here we provide results for the remaining nine tasks in Fig.~\ref{f_A_3}. Taken together with Fig.~\ref{f_5}, the overall trends across all tasks remain consistent with the earlier analysis.

\paragraph{Baseline Implementation}

We compare DLM with a set of representative offline MARL baselines, which fall into two major categories: value-based methods and imitation-based methods. Among value-based methods, TD3+BC~\citep{kostrikov2021offline} applies a conservative value estimation strategy by combining actor-critic learning with behavior cloning regularization, aiming to reduce overestimation and improve stability. MACQL~\citep{formanek2024dispelling} extends the idea of CQL~\citep{kumar2020conservative} to the multi-agent setting by incorporating joint action masking under the CTDE paradigm. OMIGA~\citep{wang2024offline} improves upon previous methods by incorporating global-to-local value shaping to better guide decentralized agents during training. CFCQL~\citep{shao2024counterfactual} further enhances robustness by introducing counterfactual regularization at the agent level, allowing better credit assignment under partial observability. On the imitation-based side, BC~\citep{syed2008apprenticeship} directly learns policies via supervised learning from offline action labels without any value estimation.

For LLM-based methods, SayCan~\citep{ahn2022can}’s input and DLM are largely similar, with the addition of selectable actions as skills in SayCan. The key difference lies in SayCan being based on an instruction-tuned LLM, specifically LLaMA-3.2-1B-Instruct. Gato, in contrast, differs from DLM in terms of trajectory construction and sequence modeling. While Gato constructs trajectories as $\tau^{(k)} = \, \left\{ (o_t^{(k)}, a_t^{(k)})\ \big|\ t = 1,\ldots, T^{(k)} \right\} \, $. Since Gato’s pre-trained models and training data are not publicly available, we align with its network architecture and train the model on our collected offline datasets. MADT~\citep{meng2023offline} formulates decision-making as an autoregressive sequence prediction problem and uses return-conditioning to generalize across tasks and agent configurations, but it requires environment interaction for online fine-tuning.

For implementation, we build all value-based baselines on top of the EPyMARL framework~\citep{papoudakis2020benchmarking}, which is designed for flexible MARL experimentation. For imitation-based baselines, we adapt publicly available implementations released by the original authors. Where no official code is available, we reproduce the algorithms based on their published descriptions and validate the implementations by replicating reported performance. All baseline methods are trained on our collected offline dataset to ensure consistency and fairness in comparison with DLM.

\paragraph{Implementation of DLM}

\begin{figure}
  \centering
  \centerline{\includegraphics[width=\textwidth]{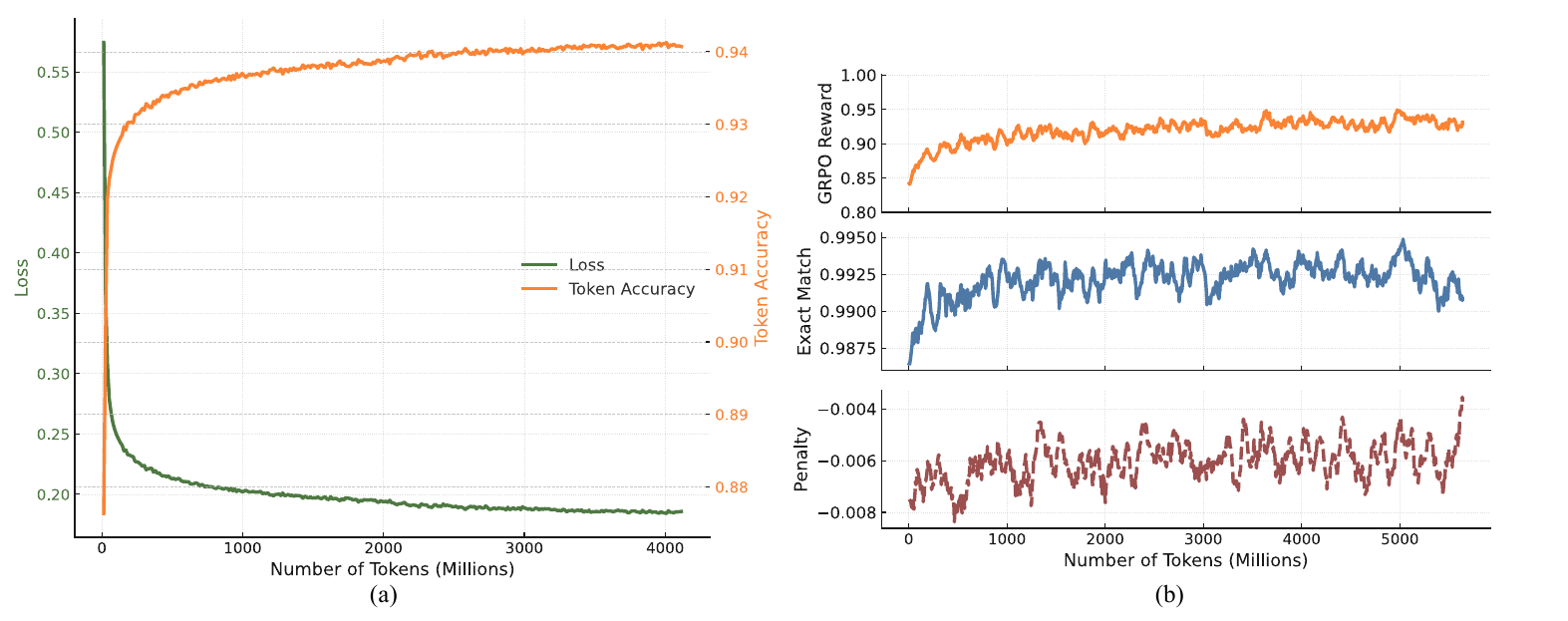}}
  \caption{Training curves of DLM. (a) DLM-SFT: cross-entropy loss and token accuracy over tokens. (b) DLM-GRPO: reward (top), exact match rate (middle), and penalty (bottom).}
  \label{f_A_2}
\end{figure}

DLM is implemented using the Hugging Face Transformers library and integrated with standard multi-agent datasets. We adopt LLaMA-3.2-1B as the pre-trained language model backbone. The training process is divided into two stages: SFT and preference-based alignment via GRPO. In the SFT stage, we perform full-parameter fine-tuning on the first half (2000 trajectories) of our collected offline dataset. All 15 SMAC tasks are mixed and fed into the model in a single training run. We choose not to use parameter-efficient methods like LoRA~\citep{hu2021lora} at this stage because our experiments show that, although LoRA can reduce training cost, it limits the model's representation capacity when dealing with diverse multi-task data. Given the small model size (1B parameters), full fine-tuning ensures sufficient capacity to fully adapt to all environments. Moreover, training on all tasks jointly avoids catastrophic forgetting that may occur if the model is fine-tuned sequentially on different tasks. In the GRPO stage, we freeze the base model from SFT and apply LoRA-based fine-tuning on the remaining half of the dataset (another 2000 trajectories). This stage focuses on reducing OOD errors by leveraging lightweight executability-based reward signals. Using LoRA here allows efficient alignment while preserving the general decision-making capability acquired during SFT. This design balances robustness and generalization and enables DLM to adapt without overwriting previously learned behaviors. Fig.~\ref{f_A_2} summarizes the learning dynamics. In subplot (a), we observe a consistent decline in SFT loss alongside a steady increase in token-level accuracy, which eventually reaches 94\%, indicating that DLM-SFT can effectively learn to reproduce behaviors from the offline trajectories across all tasks. In subplot (b), during the GRPO stage, the preference reward steadily improves. At the same time, the exact match rises suggesting that the model's outputs increasingly match preferred actions even in challenging or OOD-prone observations. The penalty term also gradually approaches zero, indicating a decreasing frequency of invalid or infeasible actions. These trends validate the effectiveness of our two-stage design in first imitating multi-agent behavior and then refining it through preference optimization.

\paragraph{Hyperparameter Settings}

All value-based baselines, including CFCQL, OMIGA, MACQL, and TD3+BC, are implemented within the EPyMARL framework. We follow the default hyperparameter settings provided in the official implementations or corresponding papers to ensure reproducibility and fairness. For example, buffer sizes are set to 5000, the learning rate is fixed at 5e-4, and $\epsilon$-greedy exploration is applied with $\epsilon$ linearly annealed from 1.0 to 0.05 over 50,000 steps. All imitation-based baselines, including MADT and BC, are trained using the same offline dataset with their original configurations where available, ensuring consistent training conditions across all methods. Key training hyperparameters for DLM are summarized in Tab.~\ref{tab:hyperparams}. All experiments are conducted on 8 NVIDIA L40 GPUs.

\begin{table}
\centering
\caption{Hyperparameters used for DLM training.}
\label{tab:hyperparams}
\begin{tabular}{lc}
\toprule
\textbf{Hyperparameter} & \textbf{Value} \\
\midrule
\multicolumn{2}{c}{\textit{DLM-SFT}} \\
\midrule
Learning rate & 2e-5 \\
Batch size & 8 \\
Gradient accumulation steps & 8 \\
Max length & 1024 tokens \\
Epochs & 2 \\
Packing & True \\
\midrule
\multicolumn{2}{c}{\textit{DLM-GRPO}} \\
\midrule
LoRA rank ($r$) & 128 \\
LoRA $\alpha$ & 256 \\
Batch size & 8 \\
Number of generations per sample & 4 \\
Epochs & 2 \\
Learning rate & 5e-5 \\
KL coefficient ($\beta$) & 0.1 \\
PPO clipping threshold ($\epsilon$) & 0.2 \\
Top-$k$ / Top-$p$ sampling & 50 / 0.95 \\
\bottomrule
\end{tabular}
\end{table}

\paragraph{Hyperparameter Tuning}

We adopt systematic strategies to tune the hyperparameters of DLM and ensure fair comparison with baseline algorithms. For all value-based methods (CFCQL, OMIGA, MACQL, and TD3+BC), we use the official hyperparameter settings from their original papers or public implementations. These configurations have been validated across the SMAC benchmark. For DLM, we perform controlled hyperparameter tuning on both the SFT and GRPO stages. Specifically, we tune the learning rate, context length, LoRA rank, KL divergence coefficient ($\beta$), PPO clipping threshold ($\epsilon$), and sampling parameters (top-$k$, top-$p$). The search process is guided by performance on a held-out subset of SMAC tasks (e.g., \texttt{3s5z}, \texttt{10m\_vs\_11m}, and \texttt{MMM2}). Considering the substantial computational cost of training across all tasks, we limit the search to a representative subset to efficiently explore the hyperparameter space. For each hyperparameter combination, we train the model under three different random seeds and select the configuration that achieves the highest average win rate across these runs.

Unlike prior methods that require task-specific tuning, DLM uses a unified set of hyperparameters across all 15 SMAC tasks. This design choice enhances generalization and prevents overfitting to any particular map. The final selected hyperparameters are summarized in Tab.~\ref{tab:hyperparams}, and all reported results are obtained using this fixed configuration without further tuning.

\paragraph{Computational Cost}

\begin{figure}
  \centering
  \includegraphics[width=\textwidth]{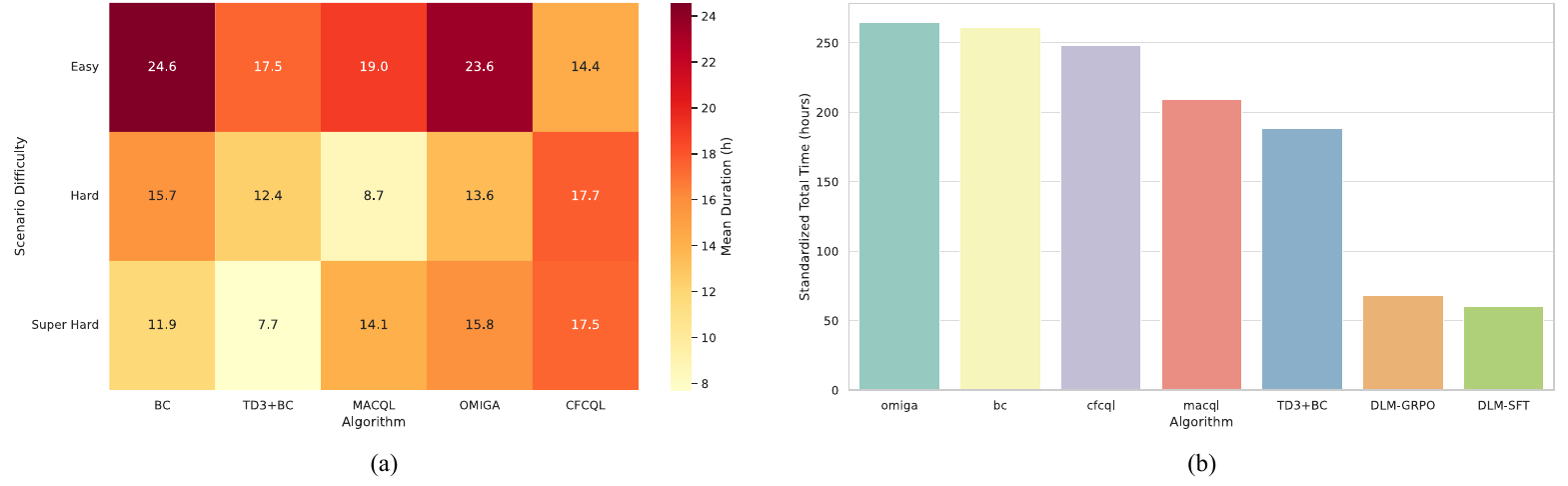}
  \caption{Computational cost comparison. (a) Average training time per difficulty level for each baseline. (b) Total standardized GPU hours across all tasks.}
  \label{f_A_4}
\end{figure}

We analyze the computational cost of all algorithms in terms of mean training duration per difficulty level and total GPU hours across all SMAC tasks. As shown in Fig.~\ref{f_A_4}(a), we observe two key trends. First, more complex algorithms, such as OMIGA and CFCQL, generally require longer training time, particularly on easy and hard tasks. Second, task difficulty tends to correlate positively with training duration, as more challenging environments typically demand longer convergence. An exception is observed with BC, which exhibits unusually high training time even on some easy tasks. A plausible explanation is that BC struggles to converge in these cases, and as a result, it often interacts with the environment until reaching the maximum episode length during each training iteration, thereby increasing the overall runtime.

In contrast, as shown in Fig.~\ref{f_A_4}(b), DLM exhibits the lowest overall computational cost among all evaluated methods, despite its two-stage training pipeline. Specifically, DLM-SFT and DLM-GRPO require approximately 60 and 70 GPU-hours respectively when trained across the full benchmark. Notably, this is significantly lower than the cumulative training cost of value-based baselines, which must be trained independently for each task. As such, the total cost reported for these baselines is obtained by summing their per-task training durations. In comparison, DLM benefits from its unified text-based formulation, enabling multi-task generalization via a single language model. This eliminates the need for repeated training or task-specific value function updates, resulting in substantial computational savings. These properties highlight DLM’s efficiency and scalability in large-scale offline MARL settings, offering a compelling balance of performance and resource efficiency.

\subsection{Additional Experimental Results}
\label{s_A_6}

\subsubsection{Ablation on GRPO Usage}

To assess the necessity of second-stage preference optimization (GRPO), we conduct an ablation study where all 4000 trajectories are used solely for SFT, omitting GRPO updates. We compare three configurations: DLM-SFT trained on 2000 trajectories, full-data SFT trained on all 4000 trajectories, and the full DLM pipeline combining SFT and GRPO. Tab.~\ref{tab:grpo_ablation} reports win rates and OOD action rates across representative SMAC tasks.

We observe that increasing the training data from 2000 to 4000 trajectories can lead to marginal improvements in some challenging environments such as \texttt{27m\_vs\_30m} and \texttt{MMM2}, where data coverage helps mitigate underfitting. For example, in \texttt{MMM2}, win rate increases from 90.8\% to 91.7\%. However, these gains are modest and come with persistently high OOD rates, indicating that naive data scaling does not eliminate OOD behavior. In easier tasks such as \texttt{2s3z}, full-data SFT provides only limited benefit and may even induce slight overfitting, as shown by the stagnating or slightly decreased win rates. In contrast, the full DLM configuration consistently improves both win rate and OOD robustness. These results confirm that data quantity alone is insufficient to address OOD generalization in multi-agent settings. This supports the analysis in Sec.~\ref{s_3_1}, that increased data alone cannot resolve OOD issues, and explicit alignment mechanisms like GRPO are essential.

\begin{table}
\centering
\caption{Comparison of win rate (\%) and OOD rate (\%) across three configurations: DLM-SFT (2000), full-data DLM-SFT (4000), and DLM (SFT+GRPO).}
\label{tab:grpo_ablation}
\begin{tabular}{lcccccc}
\toprule
\multirow{2}{*}{\textbf{Task}} & \multicolumn{3}{c}{\textbf{Win Rate (\%)}} & \multicolumn{3}{c}{\textbf{OOD Rate (\%)}} \\
\cmidrule(lr){2-4} \cmidrule(lr){5-7}
 & SFT (2000) & SFT (4000) & DLM & SFT (2000) & SFT (4000) & DLM \\
\midrule
2s3z & 98.5 & 98.4{\scriptsize ↓} & \textbf{99.7}{\scriptsize ↑} & 0.00 & 0.00 & 0.00 \\
3s5z\_vs\_3s6z & 84.2 & 81.2{\scriptsize ↓} & \textbf{89.5}{\scriptsize ↑} & 0.23 & 0.19{\scriptsize ↓} & \textbf{0.02}{\scriptsize ↓} \\
27m\_vs\_30m & 80.1 & 81.9{\scriptsize ↑} & \textbf{84.6}{\scriptsize ↑} & 2.65 & 2.41{\scriptsize ↓} & \textbf{0.78}{\scriptsize ↓} \\
MMM2 & 90.8 & 91.7{\scriptsize ↑} & \textbf{97.3}{\scriptsize ↑} & 0.70 & 0.58{\scriptsize ↓} & \textbf{0.10}{\scriptsize ↓} \\
1c3s5z & 97.7 & \textbf{99.0}{\scriptsize ↑} & 98.8{\scriptsize ↑} & 2.69 & 2.23{\scriptsize ↓} & \textbf{0.59}{\scriptsize ↓} \\
\bottomrule
\end{tabular}
\end{table}

\subsubsection{Ablation on OOD Filtering}

We further examine the impact of executability-based filtering before applying GRPO. In the ablation setting, GRPO is directly applied to all remaining trajectories without removing OOD-prone samples identified during the SFT stage. Empirically, we observe significant training instability: unlike the clean and steadily improving curves shown in Fig.~\ref{f_A_2}(b), reward signals, exact match rates, and penalties fluctuate erratically throughout training. This behavior suggests that the model struggles to converge when exposed to a mixture of already-correct and severely misaligned samples. Furthermore, in tasks where DLM-SFT already achieves high performance (e.g., \texttt{2s3z}), the absence of filtering introduces noisy gradient updates that degrade accuracy and increase training time. These results underscore the necessity of targeted optimization and validate the role of OOD filtering in enabling stable and efficient GRPO alignment.

\subsubsection{Scalability to Additional Benchmarks}

To further evaluate the scalability of DLM beyond SMAC, we expand training and testing to additional multi-agent benchmarks. Following the dataset construction methodology in Sec.~\ref{s_3_1}, we collect dialogue-style offline datasets specifically from the LBF:11$\times$11-6p-4f task~\citep{papoudakis2020benchmarking}. Based on this dataset, we train a DLM following the methodology described in this paper. Tab.~\ref{t_A_scalability_train} summarizes the results, where DLM-GRPO consistently outperforms or matches strong baselines across the benchmark.

\begin{table}[ht]
  \centering
  \caption{Performance on additional training benchmarks.}
  \label{t_A_scalability_train}
  \resizebox{\linewidth}{!}{
  \begin{tabular}{lcccccccc}
    \toprule
    \textbf{Task} & \textbf{DLM-GRPO} & \textbf{DLM-SFT} & \textbf{MADT} & \textbf{BC} & \textbf{TD3+BC} & \textbf{MACQL} & \textbf{OMIGA} & \textbf{CFCQL} \\
    \midrule
    LBF:11$\times$11-6p-4f       & $0.96 \pm 0.02$ & $0.91 \pm 0.03$ & $0.85 \pm 0.09$ & $0.28 \pm 0.06$ & $0.30 \pm 0.05$ & $0.69 \pm 0.08$ & $0.85 \pm 0.07$ & $0.77 \pm 0.09$ \\
    \bottomrule
  \end{tabular}}
\end{table}

Beyond training benchmarks, we also assess zero-shot transfer to additional tasks. On LBF:20$\times$20-10p-6f, DLM-GRPO achieves a win rate of $0.69 \pm 0.05$. These results indicate that DLM maintains strong generalization across structurally diverse environments without task-specific adaptation.

\end{document}